\documentclass[a4paper,10.9pt]{article}
\usepackage{bbding}
\usepackage{}
\usepackage{amsmath}
\usepackage{mathrsfs}
\usepackage{graphicx, subfigure}
\usepackage{wasysym}
\usepackage{color}
\usepackage{bm}
\usepackage{pifont}
\usepackage{txfonts}
\usepackage{graphicx}
\usepackage{amsfonts}
\usepackage{amssymb}
\usepackage{mathrsfs,psfrag,eepic,epsfig}
\makeatletter \@addtoreset{equation}{section}

\makeatletter
\newfont{\footsc}{cmcsc10 at 8truept}
\newfont{\footbf}{cmbx10 at 8truept}
\newfont{\footrm}{cmr10 at 10truept}
\title{\bf{Inverse scattering transform of an extended nonlinear Schr\"{o}dinger equation with nonzero boundary conditions
and its multisoliton solutions} \footnote{This work is supported by the National Natural Science Foundation of China under Grant No.11871180.\protect\\
\hspace*{3ex} $^{*}$Corresponding authors.\protect\\
\hspace*{3ex} E-mail address: xiubinwang@163.com, wangxiubin@hit.edu.cn (X.B. Wang), bohan@hit.edu.cn (B. Han)}}

\author{Xiu-Bin Wang and Bo Han$^{*}$ \\
\small Department of Mathematics, Harbin Institute of Technology, Harbin 150001,
People's Republic of China}

\date{}
\begin{document}
\maketitle

\noindent{\large \bf Abstract:}
Under investigation in this work is an extended nonlinear Schr\"{o}dinger equation with nonzero boundary conditions,
which can model the propagation of waves in dispersive media.
Firstly, a matrix Riemann-Hilbert problem for the equation with nonzero boundary conditions at infinity is systematically discussed.
Then the inverse problems are solved through the investigation of the matrix Riemann-Hilbert problem.
Therefore, the general solutions for the potentials, and explicit expressions
for the reflection-less potentials are presented.
Furthermore, we construct the simple-pole and double-pole solutions for the equation.
Finally, the remarkable characteristics of these solutions are graphically discussed.
Our results should be useful to enrich and explain the related nonlinear wave phenomena in nonlinear fields.
\\
%{\bf PACS numbers:} 02.30.Jr, 02.30.Ik, 05.45.Yv.\\

%\begin{center}
%(Some figures in this article are in colour only in the electronic version)
%\end{center}

\section{Introduction}

The standard nonlinear Schr\"{o}dinger equation (NLS)
\begin{equation}\label{NLS1}
iq_{t}+q_{xx}+2\sigma|q|^2q=0,~~\sigma=\pm1
\end{equation}
and its variants play a key role in various fields of nonlinear science
such as plasma physics \cite{HB-1993}, nonlinear optics \cite{GP-2012,YS-2013,BA-2013}, Bose-Einstein condensates \cite{PL-2016}, deep ocean \cite{AR-2009},
even finance \cite{YZY-2010} and etc \cite{yjk-1}-\cite{mg-1}.
Since it has been checked that most NLS-type equations are completely integrable,
they have rich mathematical structure. Therefore,
various research on NLS-type equations are still popular.
In particular, the investigation of exact solutions for the NLS-type equations has become
more and more attractive.
Up to now, it is still a meaningful subject to discuss soliton solutions of NLS-type equations.
Some distinct approaches have been given to derive soliton solutions of nonlinear integrable systems such as
Darboux transformation \cite{VB-1991}, Inverse scattering transform (IST) \cite{MJ-1991,jmaa-1991}, Hirota bilinear method \cite{RH-2004,wxm-2018},
Riemann-Hilbert method \cite{jky-1,mwx-1,mwx-2} and so on.
Among those methods,
it is worthy to point out that the IST is one of the most powerful tools to seek the soliton solutions of NLS-type equations.
The IST was first discussed to solve exactly the famous Korteweg-de Vries (KdV) equation with a Lax pair \cite{GS-1967}.
After that, the IST has been used to consider more and more physically important nonlinear wave equations
with their Lax pairs \cite{MJ-1981,LD-1987}.
Recently, the IST of integrable equations with NZBCs has been paid much attention to based on the solutions of the
related Riemann-Hilbert problems \cite{BP-2018}-\cite{MP-2014}.

There is no doubt that equation \eqref{NLS1} has extensive physical applications.
However, to well describe other important
types of nonlinear physical phenomena, it is necessary to go beyond the standard NLS description.
One prime research is to add higher-order terms and/or dissipative terms to the
NLS equation \eqref{NLS1} to accurately model extreme wave events in some nonlinear systems.
In addition, it is also important to preserve the integrability in the modifications of the equation \eqref{NLS1},
which has to require a subtle balance between existing terms and the newly added terms according to the physical considerations.
Along this idea, an extended nonlinear Schr\"{o}dinger equation (also called the Kundu-NLS equation) \cite{AK-1984}-\cite{Chao-2019}
was proposed and studied, whose form is
\begin{equation}\label{gtc-NLS}
iu_{t}+u_{xx}+2\epsilon^2\left(|u|^2-q_{0}^2\right)u-\left(\gamma_{t}+\gamma_{x}^2-i\gamma_{xx}\right)u+2i\gamma_{x}u_{x}=0,
\end{equation}
where $\gamma(x,t)$ is an arbitrary gauge function, and $\epsilon$ is a real parameter.

In this work, we mainly focus on the initial value problem for the Kundu-NLS equation \eqref{gtc-NLS} with the following nonzero boundary conditions
(NZBCs) as infinity
\begin{equation}\label{NZBC1}
\lim_{x\rightarrow\infty}u(x,t)=u_{\pm},
\end{equation}
with $|u_{\pm}|=q_{0}\neq 0$.
The Kundu-NLS equation \eqref{gtc-NLS} is completely integrable, its Lax pair is
\begin{equation}\label{RHP-1}
\left\{ \begin{aligned}
&\phi_{x}=U\phi,~~U=-ik\sigma_{3}+\epsilon Q\notag\\
&\phi_{t}=V\phi,~~V=-2ik^2\sigma_{3}+\widetilde{Q},
             \end{aligned} \right.
\end{equation}
where $\sigma_{3}=\mbox{diag}\{1,-1\}$, and
\begin{align}\label{RHP-2}
Q=\left(
    \begin{array}{cc}
      0 &  ue^{i\gamma}\\
      -u^{*}e^{-i\gamma}  & 0 \\
    \end{array}
  \right),~~
\widetilde{Q}=2\epsilon kQ+\left(
               \begin{array}{cc}
                 i\epsilon^2\left(|u|^2-q_{0}^2\right) & i \epsilon\left(ue^{i\gamma}\right)_{x} \\
                 i\epsilon\left(u^{*}e^{-i\gamma}\right)_{x} & -i\epsilon^2\left(|u|^2-q_{0}^2\right) \\
               \end{array}
             \right),
\end{align}
with $u^{*}(x,t)$ standing for the complex conjugate of $u(x,t)$, and $k$ is a constant spectral parameter.
Particularly if $\gamma=\sigma\int^{x}|u(y)|^2dy$,
it may be reduced to the Kundu-Eckhaus equation \cite{gxg-1999,wds-2019}, which can be used to model the propagation of waves in dispersive media.
In contrast to the NLS equation \eqref{NLS1}, Eq.\eqref{gtc-NLS} possesses extra
higher-order nonlinear terms if we choose $\gamma$ to be a function of $|u|^2$.
Furthermore, the phase of the Kundu-NLS equation \eqref{gtc-NLS} is different from that of the NLS equation \eqref{NLS1}
because of the appearance of $\gamma$.

It is well-known that the IST is a powerful approach to
derive soliton solutions.
However, the research in this work, to our knowledge, has not been conducted so far.
The chief idea of the present article is to study the multi-soliton solutions of the Kudun-NLS equation \eqref{gtc-NLS} with NZBCs
by utilizing inverse scattering transform.
Additionally, some graphic analysis
are presented to help readers to understand the propagation phenomena of these solutions.

The layout of this work is as follows.
In section 2, we investigate the direct scattering problem for the Kundu-NLS equation \eqref{gtc-NLS} with NZBCs \eqref{NZBC1}
starting from its spectral problem.
In Section 3, we investigate
the Kundu-NLS equation \eqref{gtc-NLS} with NZBCs \eqref{NZBC1} such that its simple-pole solution is found via solving the
matrix Riemann-Hilbert problems with the reflectionless potentials, and their trace formulae and theta condition are also presented.
Following a similar way, in section 4 we investigate
the Kundu-NLS equation \eqref{gtc-NLS} with NZBCs \eqref{NZBC1} such that its double-pole solutions are found via solving the
matrix Riemann-Hilbert problems.
Additionally, the dynamic behaviors of the soliton solutions are discussed
with some graphics.

%The main result of our work is the following.

\section{Direct scattering with NZBCs}
Here we study the direct scattering problem for the Kundu-NLS equation \eqref{gtc-NLS} with NZBCs \eqref{NZBC1}
starting from its spectral problem.

$~~~~~~$
{\rotatebox{0}{\includegraphics[width=5.0cm,height=5.0cm,angle=0]{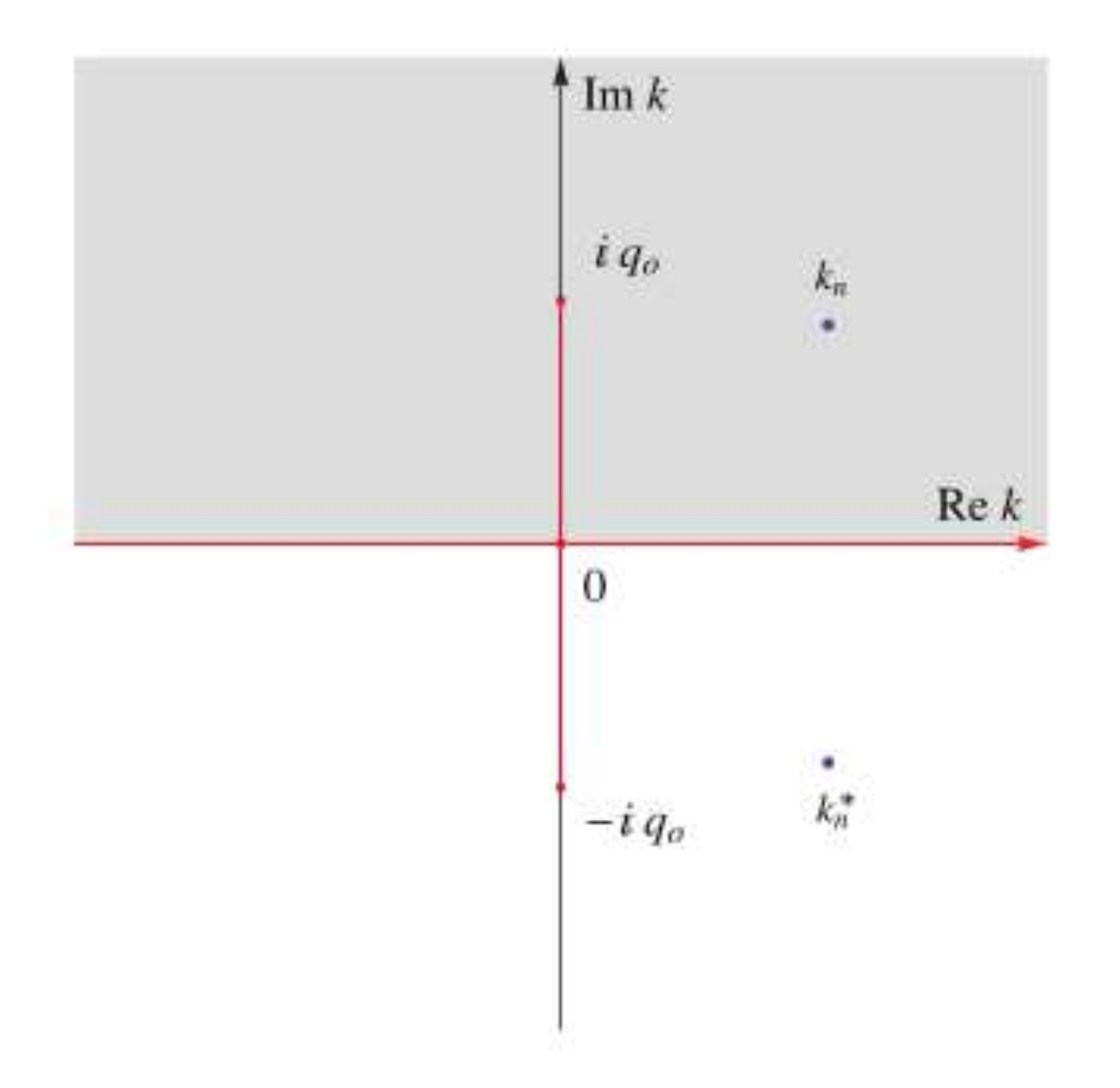}}}
~~~~~~~~~~~~~~~
{\rotatebox{0}{\includegraphics[width=5.0cm,height=5.0cm,angle=0]{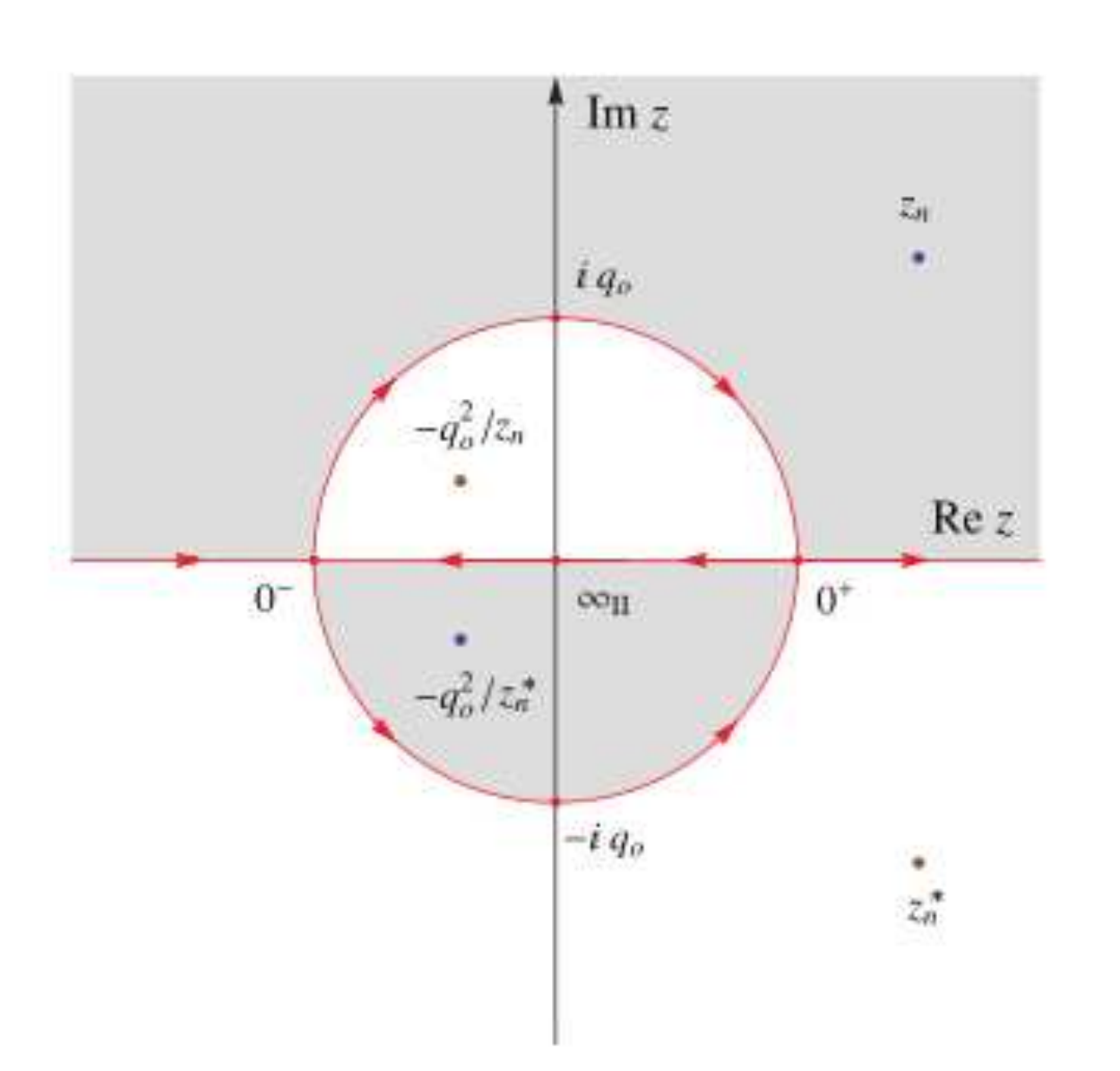}}}

$\qquad\qquad\qquad\qquad\textbf{(a)}
\qquad\qquad\qquad\qquad\qquad\qquad\qquad\qquad\qquad\textbf{(b)}
$\\
\noindent {\small \textbf{Figure 1.} The grey (white) region for $\mbox{Im}\lambda>0 (\mbox{Im}\lambda<0)$ in distinct spectral planes of the Lax pair with NZBCs. $\textbf{(a)}$: the first sheet of the Riemann surface, showing the discrete spectrum; $\textbf{(b)}$: the complex z-plane,
showing the discrete spectrum [zeros of $s_{11}(z)$ (blue) in the grey region and those of
$s_{22}(z)$ (red) in the white region], and the orientation of the jump contours for the related Riemann-Hilbert problem. \\}

Firstly we refer to the first expression in \eqref{RHP-1} as the scattering problem of equation \eqref{NZBC1}.
Asymptotically as $x\rightarrow\pm\infty$, the scattering problem reaches to
\begin{equation}\label{Lax-4}
\phi_{x}=U_{\pm}\phi,~~U_{\pm}=\lim_{x\rightarrow\pm\infty}U=ik\sigma_{3}+\epsilon Q_{\pm},
\end{equation}
and
\begin{equation}\label{Lax-41}
q=\epsilon ue^{i\gamma},~~Q_{\pm}=\lim_{x\rightarrow\pm\infty}Q(x,t)=\left(
                                                                                            \begin{array}{cc}
                                                                                              0 & q_{\pm} \\
                                                                                              -q_{\pm}^{*} & 0 \\
                                                                                            \end{array}
                                                                                          \right).
\end{equation}
Thus the fundamental matrix solutions of Eq.\eqref{Lax-4} is determined by
\begin{equation}\label{Lax-5}
\phi_{bg}^{f}(x,t,k)=\left\{ \begin{aligned}
&E^{f}_{\pm}(k)e^{i\theta(x,t,k)\sigma_{3}},~~~~k\neq \pm iq_{0},\\
&I+(x-2kt)U_{\pm},~~k=\pm iq_{0},
             \end{aligned} \right.
\end{equation}
where $I$ is a $2\times2$ unit matrix, and
\begin{equation}\label{Lax-6}
E^{f}_{\pm}(k)=\left(
                 \begin{array}{cc}
                   1 & \frac{iq_{\pm}}{k+\lambda} \\
                   \frac{iq^{*}_{\pm}}{k+\lambda} & 1 \\
                 \end{array}
               \right),~~\theta(x,t,k)=\lambda(k)\left(x-2kt\right),
\end{equation}
with
\begin{equation}\label{Lax-7}
\lambda^2=k^2+q_{0}^2.
\end{equation}
To further analyze the analyticity of the Jost solutions of the Lax pair \eqref{RHP-1},
we have to present the regions of $\mbox{Im} \lambda(k)>0 (<0)$ in the function $\theta(x,t,k)$ (also see \cite{BP-2014}).
If $q_{0}\neq0$, i.e., the boundary conditions are NZBCs, $\lambda(k)$ satisfying \eqref{Lax-7}
in the complex plane is a doubly branched function of $k$ with two branch points being $k\neq\pm iq_{0}$
and the branch cut being the segment $iq_{0}[-1,1]$.
Assume $k\pm iq_{0}=r_{\pm}e^{i\theta_{\pm}+2im_{\pm}\pi}(r_{\pm}>0,\theta_{\pm}\in[-\pi/2,3\pi/2], m\in\mathbb{Z})$.
Then two single-valued analytical branches of the complex k-plane are given by Sheet-I: $\lambda_{I}(k)=\sqrt{r_{+}r_{-}}e^{i(\theta_{+}+\theta_{-})/2}$
and Sheet-II: $\lambda_{II}(k)=-\lambda_{I}(k)$.
For this purpose, one can give a uniformization variable $z$ provided by the conformal mapping: $z=k+\lambda$,
whose inverse mapping reads
\begin{equation}\label{Lax-8}
k(z)=\frac{1}{2}\left(z-\frac{q_{0}^2}{z}\right),~~\lambda(z)=\frac{1}{2}\left(z+\frac{q_{0}^2}{z}\right).
\end{equation}
The mappings between the two-sheeted Riemann k-surface and complex z-plane are displayed in Fig.1.
If $q_{0}=0$, the NZBCs can be reduced to the ZBC. %where $\lambda(k)$ satisfying $\lambda^2(k)=k^2$
%in the complex plane is also a doubly branched function of $k$ with two branch points being $k=0$ and $k=\infty$
%and the branch cut chosen as the segment $[0,+\infty]$.
%Thus, we have $\lambda_{I}=k$ and $\lambda_{II}=-k$ for Arg($\lambda_{I}\in[0,2\pi)$).
%We have $\mbox{Im}\lambda_{I}(k)=\mbox{Im} k>0(<0)$, i.e.,
%the upper-half (lower-half) plane of k-plane, and  $\mbox{Im}\lambda_{II}(k)=\mbox{Im} k>0(<0)$,
%i.e., the lower-half (upper-half) plane of k-plane.
%Moreover, $\lambda(k)\in\mathbb{R}$ only for $k\in\mathbb{R}$.

Take $\mathbb{A}=iq_{0}[-1,1]$ with %$\mathbb{A}^{-}=iq_{0}[-1,0]$ and $\mathbb{A}^{+}=iq_{0}[0,1]$,
$C_{0}=\{z\in\mathbb{C}:|z|=q_{0}\}$, and
\begin{equation}\label{Lax-9}
D_{+}^{f}=\left\{z\in\mathbb{C}:\left(|z|^2-q_{0}^2\right)\mbox{Im}z>0\right\},~~D_{-}^{f}=\left\{z\in\mathbb{C}:\left(|z|^2-q_{0}^2\right)\mbox{Im}z<0\right\}.
\end{equation}
The continuous spectrum of $U_{\pm}=\lim\limits_{x\rightarrow\pm\infty}U$ represents the set of all values of z satisfying
$\lambda(z)\in\mathbb{R}$, i.e., $z\in\Sigma^{f}=\mathbb{R}\bigcup C_{0}$,
which are the jump contours for the related RHP.
Since $V_{\pm}=-2kUX_{\pm}$, i.e, $[U_{\pm},V_{\pm}]=0$.
As a result, similar to \cite{BP-2014},
one can find simultaneously the Jost
solutions $\phi_{\pm}(x,t,z)$ of both parts of \eqref{RHP-1} satisfying the boundary conditions
\begin{equation}\label{Lax-10}
\phi_{\pm}(x,t,z)=E_{\pm}^{f}(z)e^{i\theta(x,t,z)\sigma_{3}}+O(1),~~\forall z\in\Sigma^{f},~~x\rightarrow\pm\infty.
\end{equation}
It follows from $\phi_{x}=U_{\pm}\phi+\Delta Q_{\pm}\phi$, $\Delta Q_{\pm}(x,t)=Q(x,t)-Q_{\pm}$
with $Q_{\pm}=\lim\limits_{x\rightarrow\pm\infty}Q$ that the modified Jost solutions
\begin{equation}\label{Lax-11}
\mu_{\pm}(x,t,z)=\phi_{\pm}(x,t,z)e^{-i\theta(x,t,z)\sigma_{3}}\rightarrow E_{\pm}^{f}(z),~~x\rightarrow\pm\infty,
\end{equation}
possess the following expressions
\begin{align}\label{Lax-12}
\mu_{\pm}=
\left\{ \begin{aligned}
&E_{\pm}^{f}(z)\left\{I+\int_{\pm\infty}^{x}e^{i\lambda(x-y)\widehat{\sigma}_{3}}\left[(E_{\pm}^{f}(z))^{-1}\Delta Q_{\pm}(y,t)\mu_{\pm}(y,t,z)\right]dy\right\},\\
&~~~~~~~~~z\neq\pm iq_{0},~~q-q_{\pm}\in L^{1}\left(\mathbb{R}^{\pm}\right),\\
&E_{\pm}^{f}(z)+\int_{\pm\infty}^{x}\left[I+(x-y)\left(Q_{\pm}\mp q_{0}\sigma_{3}\right)\right]\Delta Q_{\pm}(y,t)\mu_{\pm}(y,t,z)dy,\\
&~~~~~~~~~z=\pm iq_{0},~~\left(1+|x|\right)\left(q-q_{\pm}\right)\in L^{1}\left(\mathbb{R}^{\pm}\right),
             \end{aligned} \right.
\end{align}
where $e^{\widehat{\sigma}_{3}}A:=e^{\sigma_{3}}Ae^{-\sigma_{3}}$.

Let $\Sigma_{0}^{f}:=\Sigma^{f}\setminus\{\pm iq_{0}\}$, $\mu_{\pm}(x,t,z)=(\mu_{\pm,1},\mu_{\pm,2})$,
and $\phi_{\pm}(x,t,z)=(\phi_{\pm1},\phi_{\pm2})$.
Because they include $e^{\pm i(x-y)}$,
therefore from the properties of these functions in distinct domains and the definition \eqref{Lax-12}
of $\mu_{\pm}(x,t,z)$ as well as the relation \eqref{Lax-11} between $\mu_{\pm}(x,t,z)$ and $\phi(x,t,z)$,
the following proposition is easily established (also see \cite{BP-2014} for the more details).\\

\noindent
\textbf{Proposition 2.1.} Assume $q-q_{\pm}\in L^{1}(\mathbb{R}^{\pm})$, the modified forms $\mu_{\pm2}(x,t,z)$
and the Jost functions $\phi_{\pm2}$ presented by \eqref{Lax-11} and \eqref{Lax-12}
have unique solutions in $\Sigma_{0}^{f}$.
Additionally, $\mu_{+1}(x,t,z)$, $\mu_{-2}(x,t,z)$, $\phi_{+1}(x,t,z)$ and $\phi_{-2}(x,t,z)$
can be continuously to $D_{+}^{f}\bigcup\Sigma_{0}^{f}$, and analytically extended to $D_{+}^{f}$,,
On the contrary, $\mu_{-1}(x,t,z)$, $\mu_{+2}(x,t,z)$, $\phi_{-1}(x,t,z)$ and $\phi_{+2}(x,t,z)$
can be continuously extended to $D_{-}^{f}\bigcup\Sigma_{0}^{f}$, and analytically extended to $D_{-}^{f}$.

As $\mbox{tr}U(x,t,z)=\mbox{tr}V(x,t,z)=0$ in \eqref{RHP-1} reaches to $(\det\phi_{\pm})_{x}=(\det\phi_{\pm})_{t}=0$,
one can thus see
\begin{equation}\label{Lax-13}
\det \phi_{\pm}=\lim_{x\rightarrow\pm\infty}\mu_{\pm}=\det E_{\pm}^{f}(z)=\gamma_{f}(z)=1+q_{0}^2/z^2\neq0,~~z\neq\pm iq_{0},
\end{equation}
on the basic of Liouville's formula.
Since $\phi_{\pm}(x,t,z)$ are both fundamental matrix solutions of the spectral problem \eqref{RHP-1},
as a consequence, we have
\begin{equation}\label{Lax-14}
\phi_{+}(x,t,z)=\phi_{-}(x,t,z)S(z),~~z\in\Sigma_{0}^{f},
\end{equation}
where $S(z)=(s_{ij}(z))_{2\times2}$ with $s_{ij}(z)$ being called scattering coefficients.
In view of \eqref{Lax-14}, we have
\begin{equation}\label{Lax-15}
\left\{ \begin{aligned}
&s_{11}(z)=\gamma_{f}^{-1}(z)|\phi_{+1}(x,t,z),\phi_{-2}(x,t,z)|,~~s_{12}(z)=\gamma_{f}^{-1}(z)|\phi_{+1}(x,t,z),\phi_{+2}(x,t,z)|,\\
&s_{21}(z)=\gamma_{f}^{+2}(z)|\phi_{+1}(x,t,z),\phi_{-2}(x,t,z)|,~~s_{22}(z)=\gamma_{f}^{-1}(z)|\phi_{+1}(x,t,z),\phi_{+1}(x,t,z)|,
             \end{aligned} \right.
\end{equation}
and $\det S(z)=1$.

From Proposition 2.1, it is easily found that the scattering coefficients $s_{11}(z)$ and $s_{22}(z)$ in $z\in\Sigma_{0}^{f}$
can be continuously extended to
$D_{+}^{f}\bigcup\Sigma_{0}^{f}$ and $D_{-}^{f}\bigcup\Sigma_{0}^{f}$, and analytically extended to $D_{+}^{f}$ and $D_{-}^{f}$, respectively.

To tackle the matrix Riemann-Hilbert problem in the next section, here we assume that the potential does
not admit spectral singularities, i.e., $s_{11}(z)s_{22}(z)\neq 0$ for $z\in\Sigma^{f}$
and $S(z)$ is continuous for $z=iq_{0}$.
As a consequence, the reflection coefficients are determined by
\begin{equation}\label{Lax-16}
\rho(z)=\frac{s_{21}(z)}{s_{11}(z)},~~\hat{\rho}(z)=\frac{s_{21}(z)}{s_{22}(z)},~~\forall z\in\Sigma^{f}.
\end{equation}

\section{Inverse scattering with NZBCs: Simple pole}
To derive the discrete spectrum and residue conditions in the inverse scattering problem,
we next discuss the symmetries of the scattering matrix $S(k)$.
Similar to \cite{BP-2014}, we have $k(z)=k^{*}(z^{*})$, $k(z)=k\left(-q_{0}^2/z\right)$, $\lambda(z)=\bar{\lambda}(z^{*})$,
and $\lambda(z)=-\lambda\left(-q_{0}^2/z\right)$, therefore the symmetries of $U$, $V$, and $\theta$ yield
\begin{equation}\label{ISP-1}
\left\{ \begin{aligned}
&U(x,t,z)=\sigma_{2}U^{*}(x,t,z^{*})\sigma_{2},~~U(x,t,z)=U\left(x,t,-q_{0}^2/z\right),\\
&V(x,t,z)=\sigma_{2}\bar{V}(x,t,z^{*})\sigma_{2},~~V(x,t,z)=V\left(x,t,-q_{0}^2/z\right),\\
&\theta(x,t,z)=\theta^{*}(x,t,z^{*}),~~\theta(x,t,z)=-\theta\left(x,t;-q_{0}^2/z\right),
           \end{aligned} \right.
\end{equation}
where $\sigma_{2}=\left(
                   \begin{array}{cc}
                     0 & -i \\
                     i & 0 \\
                   \end{array}
                 \right)$.

It follows from the above-mentioned symmetries and \eqref{Lax-11} that
\begin{equation}\label{ISP-2}
\left\{ \begin{aligned}
&\phi_{\pm}(x,t,z)=\sigma_{2}\phi^{*}_{\pm}(x,t,z^{*})\sigma_{2},~~\phi_{\pm}(x,t,z)=\frac{i}{z}\phi_{\pm}\left(x,t,-q_{0}^2/z\right)\sigma_{3}Q_{\pm},\\
&\mu_{\pm}(x,t,z)=\sigma_{2}\mu^{*}_{\pm}(x,t,z^{*})\sigma_{2},~~\mu_{\pm}(x,t,z)=\frac{i}{z}\mu_{\pm}\left(x,t,-q_{0}^2/z\right)\sigma_{3}Q_{\pm}.
           \end{aligned} \right.
\end{equation}
By comparing Eq.\eqref{ISP-2} with Eq.\eqref{Lax-14}, we have
\begin{equation}\label{ISP-3}
S(z)=\sigma_{2}S^{*}(z^{*})\sigma_{2},~~S(z)=\left(\sigma_{3}Q_{-}\right)^{-1}S\left(-q_{0}^2/z\right)\sigma_{3}Q_{+},
\end{equation}
which can arrive at the symmetries between $\rho(z)$ and $\hat{\rho}(z)$ as
\begin{equation}\label{ISP-4}
\rho(z)=-\hat{\rho}^{*}(z^{*}),~~\rho(z)=\frac{q^{*}_{-}}{q_{-}}\rho\left(-q_{0}^2/z\right).
\end{equation}
The discrete spectrum of the scattering problem represents the set of all values
$z\in\mathbb{C}\setminus\Sigma^{f}$ such that they have eigenfunctions in $L^2(\mathbb{R})$.
Similar to \cite{BP-2014}, they need satisfy $s_{11}(z)=0$ for $z\in D_{+}^{f}$
and $s_{22}(z)=0$ for $z\in D_{-}^{f}$ such that the corresponding eigenfunctions are in $L^2(\mathbb{R})$
from \eqref{Lax-15} and the expression \eqref{Lax-11} of $\phi_{\pm}$.

First of all, we assume that $s_{11}(z)$ possesses $N$ simple zeros in $D_{+}^{f}\bigcap\{z\in\mathbb{C}:|z|>q_{0}, \mbox{Im}z>0\}$
provided by $z_{n}$, $n=1,2,\ldots,N$, i.e., $s_{11}(z_{n})=0$ and $s'_{11}(z_{n})\neq0(n=1,2,\ldots,N)$.
If $s_{11}(z_{n})=0$,
we have $s_{22}(z^{*}_{n})=s_{22}\left(-q_{0}^2/z_{n}\right)=s_{11}\left(-q_{0}^2/z^{*}_{n}\right)=0$,
Then the set of discrete spectrum reads
\begin{equation}\label{ISP-5}
Z^{f}=\left\{z_{n},-\frac{q_{0}^2}{z^{*}_{n}},z^{*}_{n},-\frac{q_{0}^2}{z_{n}}\right\}_{n=1}^{N},
~~s_{11}(z_{n})=0,~~z_{n}\in D_{+}^{f}\cap\left\{z\in\mathbb{C}:|z|>q_{0},\mbox{Im}z>0\right\}.
\end{equation}
Since $s_{11}(z_{0})=0$ and $s'_{11}(z_{0})\neq0$ are set for $z_{0}\in Z^{f}\bigcap D_{+}^{f}$,
then based on the first expression in \eqref{Lax-15}, a norming constant $b_{+}(z_{0})$ satisfies
\begin{equation}\label{ISP-6}
\phi_{+1}(x,t,z_{0})=b_{+}(z_{0})\phi_{-2}(x,t,z_{0}).
\end{equation}
The residue condition of $\phi_{+1}(x,t,z)/s_{11}(z)$ in $z_{0}\in Z^{f}\bigcap D_{+}^{f}$ reaches to
\begin{equation}\label{ISP-7}
\mathop{\mbox{Res}}\limits_{z=z^{*}_{0}}\left[\frac{\phi_{+1}(x,t,z)}{s_{11}(z)}\right]
=\frac{\phi_{+1}(x,t,z_{0})}{s'_{11}(z_{0})}=\frac{b_{+}(z_{0})}{s'_{11}(z_{0})}\phi_{-2}(x,t,z_{0}).
\end{equation}
Following the similar way, from $s_{22}(z^{*}_{0})=0$ and $s'_{22}(z^{*}_{0})\neq0$ for $z^{*}_{0}\in Z^{f}\bigcap D_{-}^{f}$
and the second expression in \eqref{Lax-15},
we find that a norming constant $b_{-}(z^{*}_{0})$ yields
\begin{equation}\label{ISP-8}
\phi_{-2}(x,t,z^{*}_{0})=b_{-}(z^{*}_{0})\phi_{-1}(x,t,z^{*}_{0}).
\end{equation}
The residue condition $\phi_{+2}(x,t,z)/s_{22}(z)$ in $z^{*}_{0}\in Z^{f}\bigcap D_{-}^{f}$ arrives at
\begin{equation}\label{ISP-9}
\mathop{\mbox{Res}}\limits_{z=z^{*}_{0}}\left[\frac{\phi_{+2}(x,t,z)}{s_{22}(z)}\right]=\frac{\phi_{+2}(x,t,z^{*}_{0})}{s'_{22}(z^{*}_{0})}
=\frac{b_{-}(z^{*}_{0})}{s'_{22}(z^{*}_{0})}\phi_{-1}(x,t,z^{*}_{0}).
\end{equation}
For the sake of convenience, we rewrite \eqref{ISP-7} and \eqref{ISP-9} as
\begin{equation}\label{ISP-10}
\left\{ \begin{aligned}
&\mathop{\mbox{Res}}\limits_{z=z_{0}}\left[\frac{\phi_{+1}(x,t,z)}{s_{11}(z)}\right]=
A_{+}[z_{0}]\phi_{-2}(x,t,z_{0}),~~A_{+}[z_{0}]=\frac{b_{+}(z_{0})}{s'_{11}(z_{0})},~~z_{0}\in Z^{f}\cap D_{+}^{f},\\
&\mathop{\mbox{Res}}\limits_{z=z^{*}_{0}}\left[\frac{\phi_{+1}(x,t,z)}{s_{11}(z)}\right]=
A_{+}[z^{*}_{0}]\phi_{-2}(x,t,z^{*}_{0}),~~A_{+}[z^{*}_{0}]=\frac{b_{+}(z^{*}_{0})}{s'_{11}(z^{*}_{0})},~~z^{*}_{0}\in Z^{f}\cap D_{+}^{f},
           \end{aligned} \right.
\end{equation}
which admits the following symmetries
\begin{equation}\label{ISP-11}
A_{+}[z_{0}]=-A^{*}_{-}[z^{*}_{0}],~~A_{+}[z_{0}]=\frac{z_{0}^2}{q_{-}^2}A_{-}\left[-\frac{q_{0}^2}{z_{0}}\right],
~~z_{0}\in Z^{f}\cap D_{+}^{f},
\end{equation}
in terms of symmetries \eqref{ISP-2} and \eqref{ISP-3}, which lead directly to
\begin{equation}\label{ISP-12}
A_{+}[z_{n}]=-A^{*}_{-}[z^{*}_{n}]=\frac{z_{n}^2}{q_{-}^2}A_{-}\left[-\frac{q_{0}^2}{z_{n}}\right]
=-\frac{z_{n}^2}{q_{-}^2}A^{*}_{+}\left[-\frac{q_{0}^2}{z^{*}_{n}}\right].
~~z_{n}\in Z^{f}\cap D_{+}^{f},
\end{equation}
The relation $\phi_{+}(x,t,z)=\phi_{-}(x,t,z)S(z)$ admits another form
\begin{equation}\label{ISP-13}
\left\{ \begin{aligned}
&\frac{\phi_{+1}(x,t,z)}{s_{11}(z)}=\phi_{-1}(x,t,z)+\rho(z)\phi_{-2}(x,t,z),\\
&\frac{\phi_{+2}(x,t,z)}{s_{22}(z)}=\hat{\rho}(z)\phi_{-1}(x,t,z)+\phi_{-2}(x,t,z),
           \end{aligned} \right.
\end{equation}
which yields
\begin{equation}\label{ISP-14}
\left[\phi_{-1}(x,t,z),\frac{\phi_{+2}(x,t,z)}{s_{22}(z)}\right]=\left[\frac{\phi_{+1}(x,t,z)}{s_{11}(z)},\phi_{-2}(x,t,z)\right]
\left[I-J_{0}(x,t,\lambda)\right],
\end{equation}
with
\begin{equation}\label{ISP-15}
J_{0}=\left(
        \begin{array}{cc}
          0 & -\hat{\rho}(z) \\
          \rho(z) & \rho(z)\hat{\rho}(z) \\
        \end{array}
      \right).
\end{equation}
Similar to \cite{BP-2014}, the asymptotics for modified Jost solutions and scattering data satisfy
\begin{equation}\label{ISP-16}
\left\{ \begin{aligned}
&\mu_{\pm}(x,t,z)=I+O\left(\frac{1}{z}\right),~~S(z)=I+O\left(\frac{1}{z}\right),~~z\rightarrow\infty,\\
&\mu_{\pm}(x,t,z)=\frac{i}{z}\sigma_{3}Q_{\pm}+O(1),~~S(z)=\frac{q_{+}}{q_{-}}I+O(z),~~z\rightarrow 0.
           \end{aligned} \right.
\end{equation}
Based on the modified Jost functions, we have the sectionally meromorphic matrix $M(x,t,z)$ as
\begin{align}\label{ISP-17}
M(x,t,z)=
\left\{ \begin{aligned}
M^{+}(x,t,z)=&\left(\frac{\mu_{+1}(x,t,z)}{s_{11}(z)},\mu_{-2}(x,t,z)\right)\\
&=\left(\frac{\phi_{+1}(x,t,z)}{s_{11}(z)},\phi_{-2}(x,t,z)\right)e^{-i\theta(x,t,z)\sigma_{3}},~~z\in D_{+}^{f},\\
M^{-}(x,t,z)=&\left(\mu_{-1}(x,t,z),\frac{\mu_{+2}(x,t,z)}{s_{22}}\right)\\
&=\left(\phi_{-1}(x,t,z),\frac{\phi_{+2}(x,t,z)}{s_{22}}\right)e^{-i\theta(x,t,z)\sigma_{3}},~~z\in D_{-}^{f}.
           \end{aligned} \right.
\end{align}
Summarizing the above analysis, the following proposition for the Riemann-Hilbert problem holds.\\

\noindent
\textbf{Proposition 3.1}.
The matrix function $M(x,t,z)$ admits the following matrix Riemann-Hilbert problem\\

\noindent
$\bullet$ Analyticity: $M(x,t,z)$ is analytic in $\left(D_{+}^{f}\bigcup D_{-}^{f}\right)\setminus Z^{f}$;\\
$\bullet$ Jump condition: $M^{-}(x,t,z)=M^{+}(x,t,z)(I-J(x,t,z))$,~~$z\in\Sigma^{f}$ with $J(x,t,z)=e^{i\theta(x,t,z)\widehat{\sigma}_{3}}J_{0}$;\\
$\bullet$ Asymptotic behavior: $M^{\pm}(x,t,z)=I+(1/z)$ for $z\rightarrow\infty$ and $M^{\pm}=\frac{i}{z}\sigma_{3}Q_{-}+O(1)$ for $z\rightarrow0$.

To conveniently tackle the above Riemann-Hilbert problem (i.e.,Proposition 3.1), we take
\begin{equation}\label{JSP-18}
\xi_{n}=\left\{ \begin{aligned}
&z_{n},~~n=1,2,\ldots,N,\\
&-\frac{q_{0}^2}{z^{*}_{n-N}},~~n=N+1,N+2,\ldots,2N,
           \end{aligned} \right.
\end{equation}
and $\widehat{\xi}_{n}=-q_{0}^2/\xi_{n}$. Then $Z^{f}=\{\xi_{n},\widehat{\xi}_{n}\}_{n=1}^{2N}$
with $\xi_{n}\in D_{+}^{f}$ and $\widehat{\xi}_{n}\in D_{-}^{f}$.
Subtracting out the asymptotic behaviors and the simple pole contributions, i.e.,
\begin{equation}\label{JSP-19}
M_{sp}(x,t,z)=I+\frac{i}{z}\sigma_{3}Q_{-}+\sum_{n=1}^{2N}\left[\frac{\mathop{\mbox{Res}}\limits_{z=\xi_{n}}M^{+}(x,t,z)}{z-\xi_{n}}+
\frac{\mathop{\mbox{Res}}\limits_{z=\widehat{\xi}_{n}}M^{-}(x,t,z)}{z-\widehat{\xi}_{n}}\right],
\end{equation}
from both sides of the above jump condition $M^{-}=M^{+}(I-J)$, we arrive at
\begin{equation}\label{JSP-20}
M^{-}(x,t,z)-M_{sp}(x,t,z)=M^{+}(x,t,z)-M_{sp}(x,t,z)-M^{+}(x,t,z)J(x,t,z).
\end{equation}
Here $M^{\pm}(x,t,z)\rightarrow M_{sp}(x,t,z)$ is analytic in $D_{\pm}^{f}$.
In addition, the asymptotics are both $O(1/z)$ as $z\rightarrow\infty$ and $O(1)$ as $z\rightarrow0$
and $J(x,t,z)$ is $O(1/z)$ as $z\rightarrow\infty$, and $O(z)$ as $z\rightarrow 0$.
As a result, the Cauchy projectors yield
\begin{equation}\label{JSP-21}
P^{\pm}[f](z)=\frac{1}{2\pi i}\int_{\Sigma^{f}}\frac{f(\zeta)}{\zeta-(z\pm i0)}d\zeta,
\end{equation}
where the notation $z\pm i0$ is the limit taken from the left/right of $z$,
and Plemelj's formulae are presented to solve \eqref{JSP-20} and yield
\begin{equation}\label{JSP-22}
M(x,t,z)=M_{sp}(x,t,z)+\frac{1}{2\pi i}\int_{\Sigma^{f}}\frac{M^{+}(x,t,\zeta)J(x,t,\zeta)}{\zeta-z}d\zeta,~~z\in\mathbb{C}\setminus\Sigma^{f},
\end{equation}
where $\int_{\Sigma^{f}}$ represents the integral along the oriented contours displayed in Fig.1.

From \eqref{ISP-17}, we find that only its first (second) column admits a simple pole at $z=\xi_{n}(z=\widehat{\xi}_{n})$.
As a consequence, by using \eqref{Lax-11} and \eqref{ISP-10}, the residue part in \eqref{JSP-22} can be written as
\begin{align}\label{JSP-23}
\frac{\mathop{\mbox{Res}}\limits_{z=\xi_{n}} M^{+}(x,t,z)}{z-\xi_{n}}&
+\frac{\mathop{\mbox{Res}}\limits_{z=\widehat{\xi}_{n}} M^{-}(x,t,z)}{z-\widehat{\xi}_{n}}\notag\\
&=\left(\frac{A_{+}\left[\xi_{n}\right]e^{-2i\theta(x,t,\xi_{n})}}{z-\xi_{n}}\mu_{-2}\left(x,t,\xi_{n}\right),
\frac{A_{-}\left[\widehat{\xi}_{n}\right]e^{2i\theta\left(x,t,\widehat{\xi}_{n}\right)}}{z-\widehat{\xi}_{n}}\mu_{-1}\left(x,t,\widehat{\xi}_{n}\right)\right).
\end{align}

For $z=\xi_{s} (s=1,2,\ldots,2N)$, from the second column of $M(x,t,z)$ provided by \eqref{JSP-22} with \eqref{JSP-23}, we obtain
\begin{align}\label{JSP-24}
\mu_{-2}(x,t,\xi_{s})=\left(
                         \begin{array}{c}
                           \frac{iq_{-}}{\xi_{s}} \\
                           1 \\
                         \end{array}
                       \right)&+\sum_{n=1}^{2N}\frac{A_{-}\left[\widehat{\xi}_{n}\right]e^{2i\theta\left(x,t,\widehat{\xi}_{n}\right)}}
                       {\xi_{s}-\widehat{\xi}_{n}}
                       \mu_{-1}\left(x,t,\widehat{\xi}_{n}\right)\notag\\
                       &+\frac{1}{2\pi i}\int_{\Sigma^{f}}\frac{\left(M^{+}J\right)_{2}\left(x,t,\zeta\right)}{\zeta-\xi_{s}}d\zeta,~~s=1,2,\ldots,2N.
\end{align}
Furthermore, in view of \eqref{ISP-2}, we have
\begin{equation}\label{JSP-25}
\mu_{-2}(x,t,\xi_{s})=\frac{iq_{-}}{\xi_{s}}\mu_{-1}\left(x,t,\widehat{\xi}_{s}\right),~~s=1,2,\ldots,2N.
\end{equation}
Substituting \eqref{JSP-25} into \eqref{ISP-2} leads directly to
\begin{align}\label{JSP-26}
\sum_{n=1}^{2N}\left(\frac{A_{-}[\widehat{\xi}_{n}]e^{2i\theta\left(x,t,\widehat{\xi}_{n}\right)}}{\xi_{s}-\widehat{\xi}_{n}}-\frac{iq_{-}}{\xi_{s}}\delta_{sn}\right)
&\mu_{-1}\left(x,t,\widehat{\xi}_{n}\right)+\left(
                                   \begin{array}{c}
                                     \frac{iq_{-}}{\xi_{s}} \\
                                     1 \\
                                   \end{array}
                                 \right)\notag\\
                                 &+\frac{1}{2\pi i}\int_{\Sigma^{f}}\frac{\left(M^{+}J\right)_{2}\left(x,t,\zeta\right)}{\zeta-\xi_{s}}d\zeta=0,~~s=1,2,\ldots,2N,
\end{align}
where
\begin{equation*}
\delta_{sn}=\left\{ \begin{aligned}
1,~~s=n,\\
0,~~s\neq n.
           \end{aligned} \right.
\end{equation*}
System \eqref{JSP-26} admitting $2N$ equations with $2N$ unknowns $\mu_{-1}(x,t,\widehat{\xi}_{n})$
can lead to the solutions $\mu_{-1}(x,t,\widehat{\xi}_{s})$
such that one can also obtain $\mu_{-2}(x,t,\xi_{s})$ form \eqref{JSP-26}.
Consequently, plugging $\mu_{-1}(x,t,\widehat{\xi}_{s})$ and $\mu_{-2}(x,t,\xi_{s})$ into \eqref{JSP-23},
and then substitution of \eqref{JSP-23} into \eqref{JSP-22} can lead to the $M(x,t,z)$ according to the scattering data.

It follows from \eqref{JSP-23} with \eqref{JSP-22} that the asymptotic behavior of $M(x,t,z)$ reaches to
\begin{equation}\label{JSP-27}
M(x,t,z)=I+\frac{M^{(1)}(x,t)}{z}+O\left(\frac{1}{z^2}\right),~~z\rightarrow\infty,
\end{equation}
where
\begin{align}\label{JSP-28}
M^{(1)}(x,t)&=i\sigma_{3}Q+\sum_{n=1}^{2N}\left[A_{+}\left[\xi_{n}\right]e^{-2i\theta(x,t,\xi_{n})},
A_{-}\left[\widehat{\xi}_{n}\right]e^{2i\theta(x,t,\widehat{\xi}_{n})}\mu_{-1}(x,t,\widehat{\xi}_{n})\right]\notag\\
&-\frac{1}{2\pi i}\int_{\Sigma^{f}}M^{+}(x,t,\zeta)J(x,t,\zeta)d\zeta,
\end{align}
with $\mu_{-1}(x,t,\widehat{\xi}_{s})$ and $\mu_{-2}(x,t,\widehat{\xi}_{s})$ defined by \eqref{JSP-25} and \eqref{JSP-26}.

It follows from \eqref{ISP-17} that $M(x,t,z)e^{i\theta(x,t,z)\sigma_{3}}$ satisfies \eqref{RHP-1}.
Substitution of $M(x,t,z)e^{i\theta(x,t,z)\sigma_{3}}$ with \eqref{JSP-27} into the x-part of the Lax pair \eqref{RHP-1}
and then choosing the coefficients of $z^{0}$, one can arrive at the following proposition for the potential $u(x,t)$.\\

\noindent
\textbf{Proposition 3.2.} The potential with simple poles of the kundu-NLS equation \eqref{gtc-NLS} with NZBCs can be found as
\begin{equation}\label{JSP-29}
\epsilon ue^{i\gamma}=q_{-}-i\sum_{n=1}^{2N}A_{-}\left[\widehat{\xi}_{n}\right]e^{2i\theta(x,t,\widehat{\xi}_{n})}\mu_{-11}(x,t,\widehat{\xi}_{n})
+\frac{1}{2\pi}\int_{\Sigma^{f}}\left(M^{+}J\right)_{12}(x,t,\zeta)d\zeta,
\end{equation}
where $\xi_{n}=z_{n}$, $\xi_{n+N}=-q_{0}^2/z^{*}_{n-N}$, $n=1,2,\ldots,N$, $\widehat{\xi}_{n}=-q_{0}^2/\xi_{n}$, and
$\mu_{-11}(x,t,\widehat{\xi}_{n})$ are expressed by
\begin{align}\label{JSP-30}
\sum_{n=1}^{2N}\left(\frac{A_{-}\left[\widehat{\xi}_{n}\right]e^{2i\theta(x,t,\widehat{\xi}_{n})}}{\xi_{s}-\widehat{\xi}_{n}}
-\frac{iq_{-}}{\xi_{s}}\delta_{sn}\right)
&\mu_{-11}\left(x,t,\widehat{\xi}_{n}\right)+\frac{iq_{-}}{\xi_{s}}\notag\\
&+\frac{1}{2\pi i}\int_{\Sigma^{f}}\frac{\left(M^{+}J\right)_{12}\left(x,t,\zeta\right)}{\zeta-\xi_{s}}d\zeta=0,~~s=1,2,\ldots,2N,
\end{align}
which can be obtained from expression \eqref{JSP-26}.

Because $s_{11}(z)$ and $s_{22}(z)$ are analytic in $D_{+}^{f}$ and $D_{-}^{f}$, respectively,
and the discrete spectral points $\xi_{n}'$s and $\widehat{\xi}_{n}$'s are the
simple zeros of $s_{11}(z)$ and $s_{22}(z)$, respectively.
Similar to \cite{BP-2014}, the trace formulae for the kundu-NLS equation \eqref{gtc-NLS} with NZBCs can be found as
\begin{equation}\label{JSP-31}
s_{11}(z)=e^{s(z)}s_{0}(z)~~\mbox{for}~~z\in D_{+}^{f},~~
s_{22}(z)=e^{-s(z)}/s_{0}(z)~~\mbox{for}~~z\in D_{-}^{f},
\end{equation}
where
\begin{equation}\label{JSP-32}
s(z)=-\frac{1}{2\pi i}\int_{\Sigma^{f}}\frac{\log[1+\rho(\zeta)\rho^{*}(\zeta^{*})]}{\zeta-z}d\zeta,~~
s_{0}(z)=\prod_{n=1}^{N}\frac{(z-z_{n})(z+q_{0}^2/z^{*}_{n})}{(z-z^{*}_{n})(z+q_{0}^2/z_{n})}.
\end{equation}
See \cite{BP-2014} for the detailed analysis.

It follows from the limit $z\rightarrow0$ of $s_{11}(z)$ in \eqref{JSP-32} and \eqref{ISP-16}
that the following theta condition yields
\begin{equation}\label{JSP-33}
\mbox{arg}\left(\frac{q_{+}}{q_{-}}\right)=4\sum_{n=1}^{N}\mbox{arg}z_{n}+\int_{\Sigma^{f}}\frac{\log[1
+\rho(\zeta)\rho^{*}(\zeta^{*})]}{2\pi\zeta}d\zeta,~~z\rightarrow 0.
\end{equation}
Especially, for the case of the reflectionless potential, i.e., $\rho(z)=\hat{\rho}(z)=0$,
which can lead directly to $J=(0)_{2\times2}$.
Therefore, Eq.\eqref{JSP-30} arrives at
\begin{equation}\label{JSP-34}
\sum_{n=1}^{2N}\left(\frac{A_{-}[\widehat{\xi}_{n}]e^{2i\theta(x,t,\widehat{\xi}_{n})}}{\xi_{s}-\widehat{\xi}_{n}}
-\frac{iq_{-}}{\xi_{s}}\delta_{sn}\right)\mu_{-11}(x,t,\widehat{\xi}_{n})=\frac{iq_{-}}{\xi_{s}},~~s=1,2,\ldots,2N.
\end{equation}
which can be obtained for $\mu_{-11}(x,t,\widehat{\xi}_{n})$ by making use of the Cramer's rule.
Therefore, the following theorem for the potential u(x,t) in the case of simple pole holds.\\

\noindent
\textbf{Theorem 3.3.} The reflectionless potential with simple poles in the Kundu-NLS equation \eqref{gtc-NLS} with NZBCs \eqref{NZBC1} yields
\begin{equation}\label{JSP-35}
u(x,t)=\left[q_{-}+i\frac{\det\left(
                          \begin{array}{cc}
                            G & v \\
                            w^{T} & 0 \\
                          \end{array}
                        \right)
}{\det G}\right]\frac{e^{-i\gamma}}{\epsilon},
\end{equation}
where $w=(w_{j})_{2N\times1}$, $v=(v_{j})_{2N\times1}$, $G=(g_{sj})_{2N\times 2N}$, and $y=(y_{n})_{2N\times1}=G^{-1}v$
with $w_{j}=A_{-}[\widehat{\xi}_{j}]e^{2i\theta(x,t,\widehat{\xi}_{j})}$, $v_{j}=-iq_{-}/\xi_{j}$,
$g_{sj}=\frac{w_{j}}{\xi_{s}-\widehat{\xi}_{j}}+v_{s}\delta_{sj}$, and $y_{n}=\mu_{-11}(x,t,\widehat{\xi}_{n})$.

For the case of the reflectionless potential $\rho(z)=\hat{\rho}(z)=0$, the above obtained trace formulae and theta condition reduce to
\begin{equation}\label{JSP-36}
\left\{ \begin{aligned}
&s_{11}=\prod_{n=1}^{N}\frac{(z-z_{n})(z+q_{0}^2/z^{*}_{n})}{(z-z^{*}_{n})(z+q_{0}^2/z_{n})},~~\mbox{for}~~z\in D_{+}^{f},\\
&s_{22}=\prod_{n=1}^{N}\frac{(z-z^{*}_{n})(z+q_{0}^2/z_{n})}{(z-z_{n})(z+q_{0}^2/z^{*}_{n})},~~\mbox{for}~~z\in D_{-}^{f},
           \end{aligned} \right.
\end{equation}
and
\begin{equation}\label{JSP-37}
\mbox{arg}\left(\frac{q_{+}}{q_{-}}\right)=4\sum_{n=1}^{N}\mbox{arg}(z_{n}),
\end{equation}
respectively.

In what follows, we discuss the dynamic behaviors of these obtained solutions \eqref{JSP-35} of the Kundu-NLS equation \eqref{gtc-NLS}.

\noindent
\textbf{Case (1)}: Here we present a simple example for $N=1$, $\epsilon=1/2, z_{1}=3i/2, A_{+}[z_{1}]=1, q_{-}=1$ in Theorem 3.3.,
which is a time-periodic breather (also called  Kuznetsov-Ma breather) in Fig.2(a).
It follows from Fig.2(a)-(c) that when the non-zero parameter $q_{-}$ becomes smaller, the periodic behavior of the simple-pole breather
only appears in its top part, and the maximal amplitude under the background gradually decreases.
Specifically, when $q_{0}\rightarrow0$,
the breather solution of the Kundu-NLS equation \eqref{gtc-NLS} with NZBCs becomes the simple-pole bright soliton of the
Kundu-NLS equation \eqref{gtc-NLS} with ZBC (see Fig.(d)).

\noindent
\textbf{Case (2)}: Here we take $N=1, z_{1}=e^{3\pi/4}, A_{+}[z_{1}]=1, \epsilon=1/2, q_{-}=1$ in Theorem 3.3.,
we have the non-stationary breather of the Kundu-NLS equation \eqref{gtc-NLS} with NZBCs,
which is a space-periodic breather (also called Akhmediev breather) displayed in Fig.3.
When the $z_{1}$ becomes larger or smaller, this will change the direction of the wave (see Figs.3).
In addition, from Fig.3(a), we find that
the behavior of the wave is symmetrical about time $t=0$.

\noindent
\textbf{Case (3)}: Here we take $N=2, z_{1}=0.2+2i, z_{2}=1+i, \epsilon=1/2, A_{+}[z_{1}]=1,A_{+}[z_{2}]=1,q_{-}=1$ in Theorem 3.3.,
we obtain the strong interactions of simple-pole breather-breather solutions of the Kundu-NLS equation \eqref{gtc-NLS} with NZBCs
for $q_{-}=1$ (see Fig. 4(a)), $q_{-}=0.5$ (see Fig.4(b)) and $q=-0.2$ (see Fig.4(c)).
When the parameter $q_{-}$ becomes smaller,
the periodic behaviors
of the simple-pole breather-breather solutions only appear in their top parts.
Specifically, when $q_{-}\rightarrow0$, we have the strong
interactions of the simple-pole bright-bright solitons of the Kundu-NLS equation \eqref{gtc-NLS} (see Figs.4(a)-(d)).

\noindent
\textbf{Case (4)}: From Fig.5(a), we can see the interactions of simple-pole breather-breather solutions of the Kundu-NLS equation \eqref{gtc-NLS}.
When $z_{1}= 0.5i$, $z_{2}=1.5i$, one example of the breather-breather waves is shown in Fig.5(b), where both (two breather waves) have different modulation
frequencies. When $z_{1}=z_{2}=1.5i$, they appear as a first-order Akhmediev breather (see Fig.5(c)).

\noindent
\textbf{Case (5)}: If we take $z_{1}=-1/65+1.5i, z_{2}=-1/65+1.05i$ in Theorem 3.3.,
then we obtain another type of simple-pole breather-breather solutions (see Fig.6).
As shown in Fig.6, the result is a simply periodic solution.
Specifically, when $q_{-}\rightarrow0$, we have the simple-pole bright-bright solitons of the Kundu-NLS equation \eqref{gtc-NLS}.
However, the bright-bright solitons is a simply periodic solution (see Fig.6(d)).

$~~~~~~$
{\rotatebox{0}{\includegraphics[width=5.2cm,height=3.6cm,angle=0]{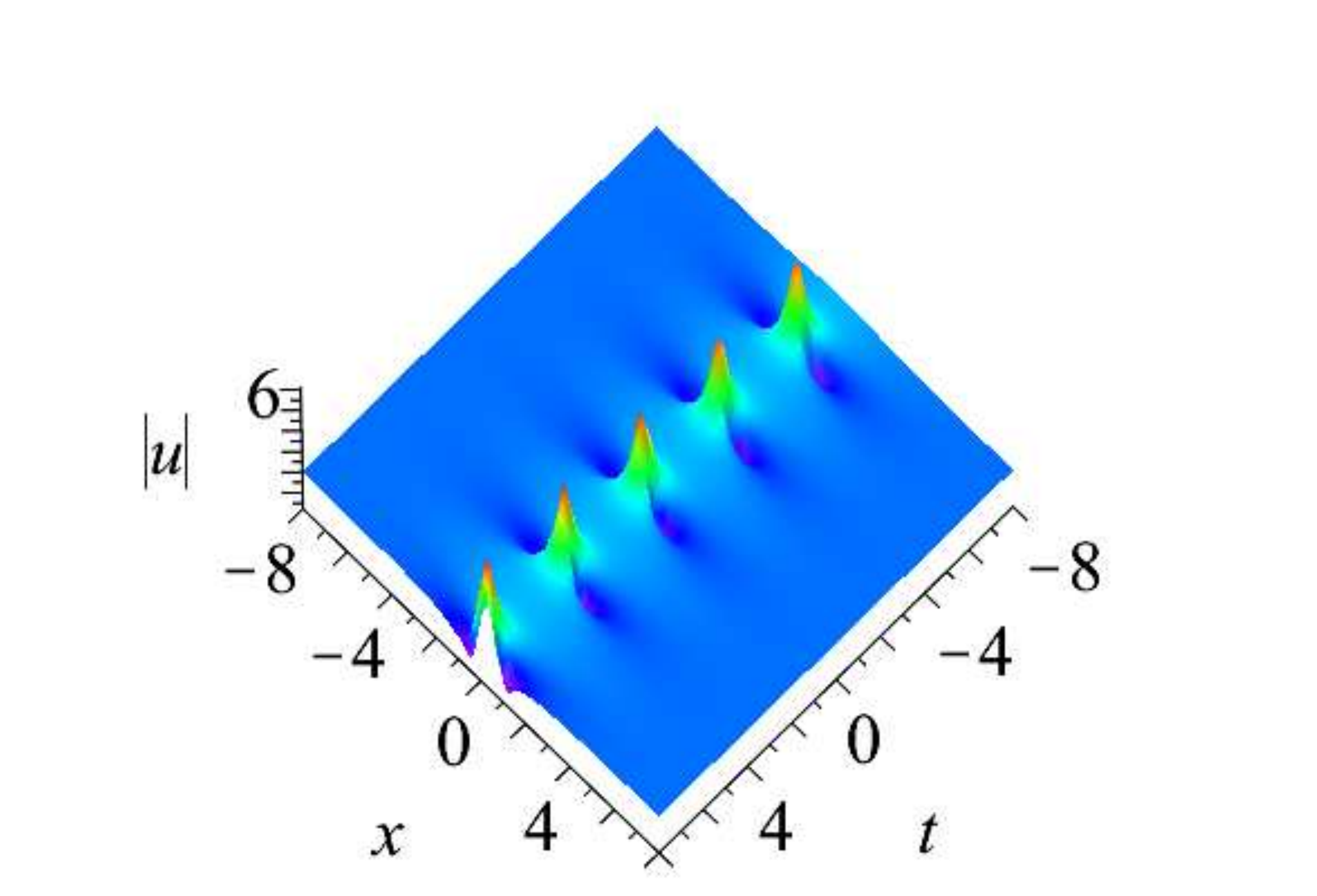}}}
~~~~~~~~~~~~~~~
{\rotatebox{0}{\includegraphics[width=5.2cm,height=3.6cm,angle=0]{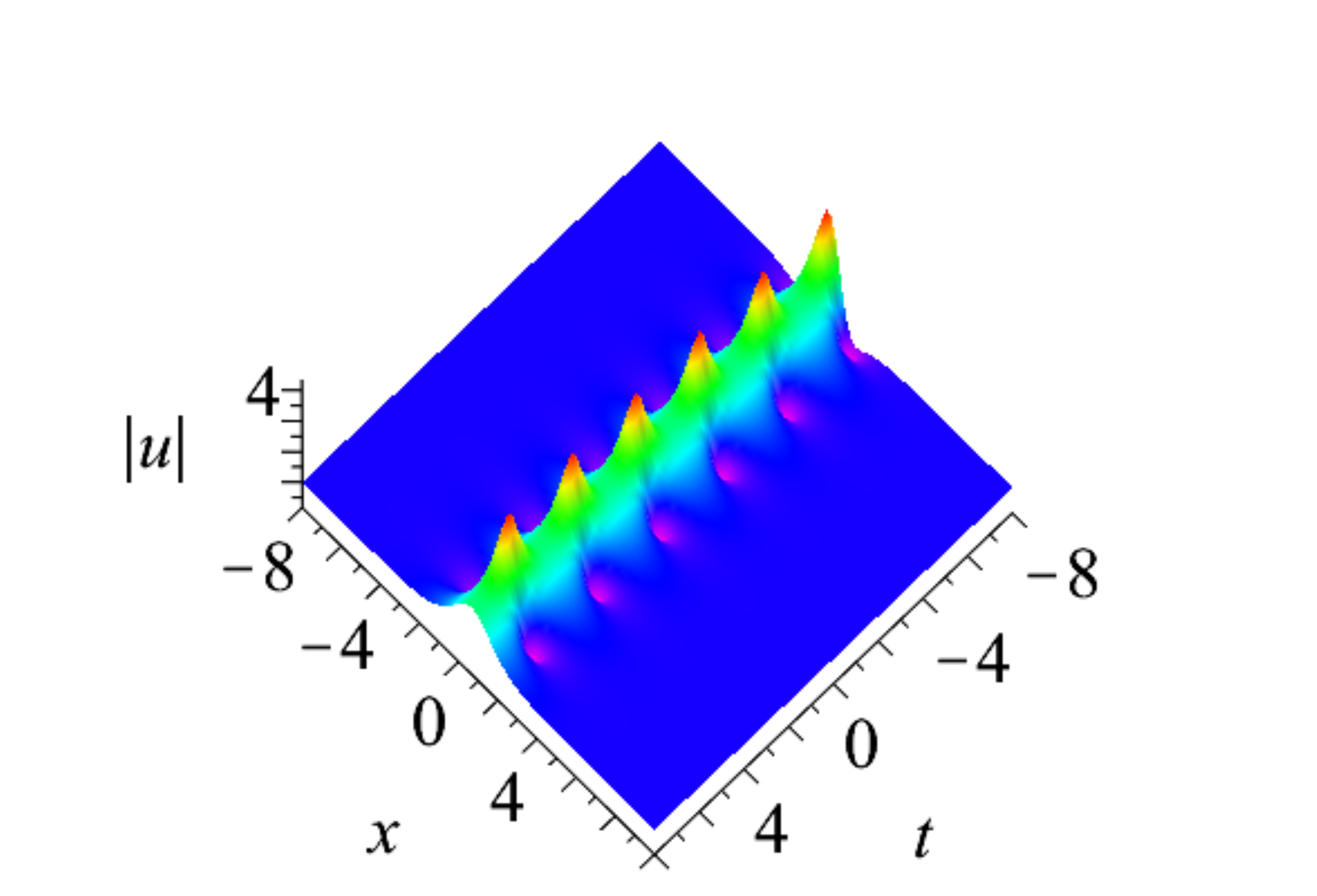}}}

$\qquad\qquad\qquad\qquad\textbf{(a)}
\qquad\qquad\qquad\qquad\qquad\qquad\qquad\qquad\qquad\textbf{(b)}
$\\

$~~~~~~$
{\rotatebox{0}{\includegraphics[width=5.2cm,height=3.6cm,angle=0]{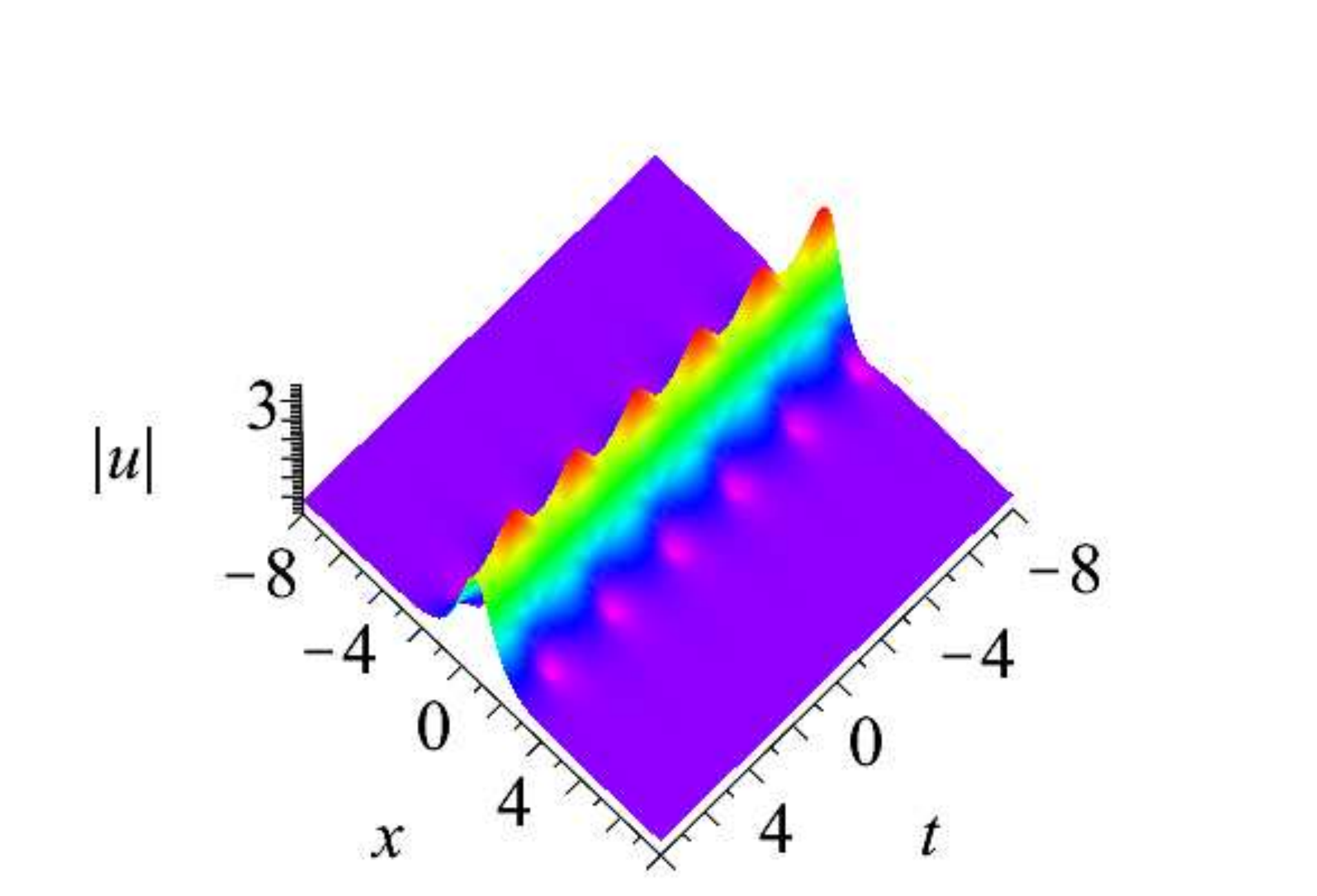}}}
~~~~~~~~~~~~~~~
{\rotatebox{0}{\includegraphics[width=5.2cm,height=3.6cm,angle=0]{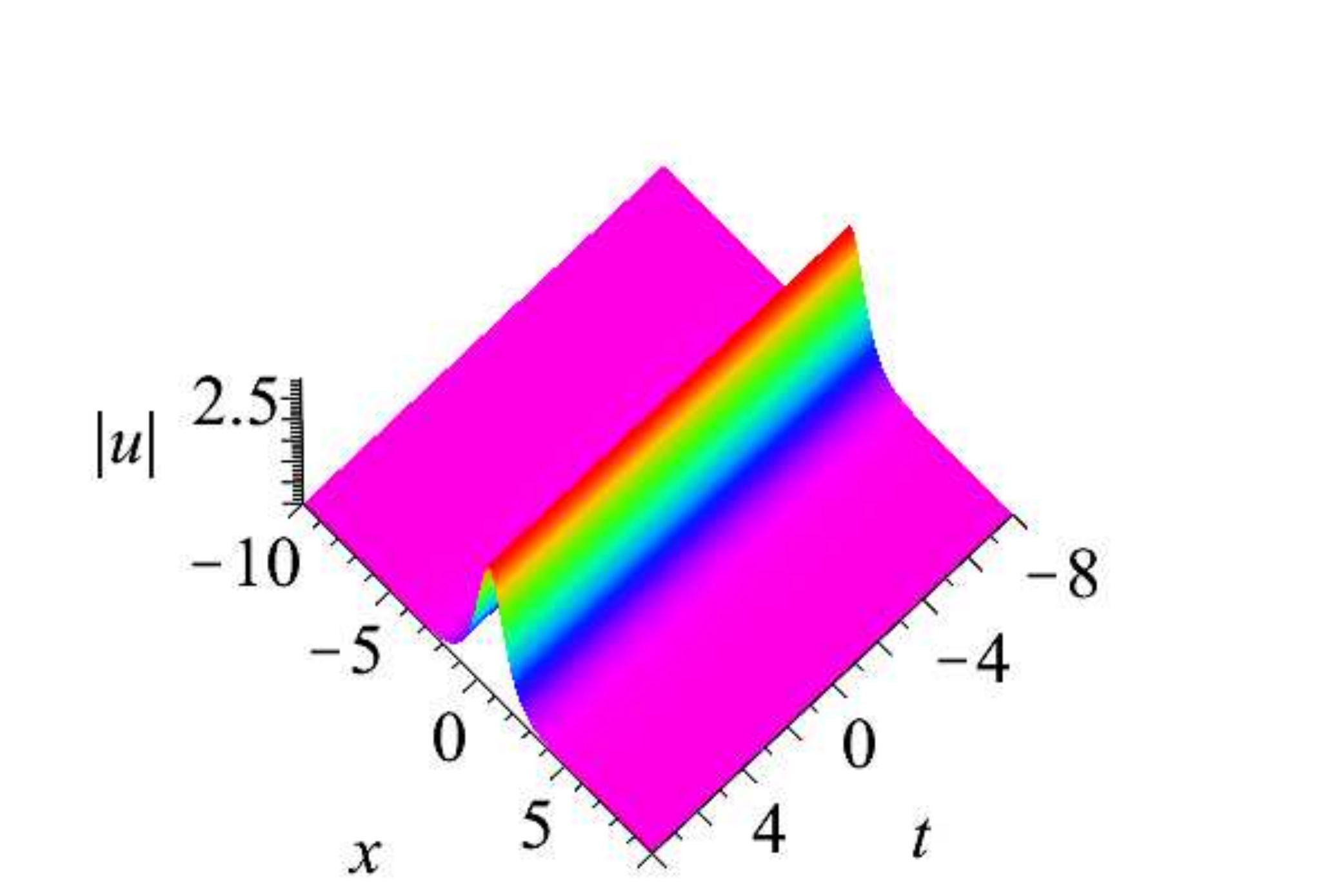}}}

$\qquad\qquad\qquad\qquad\textbf{(c)}
\qquad\qquad\qquad\qquad\qquad\qquad\qquad\qquad\qquad\textbf{(d)}
$\\
\noindent { \small \textbf{Figure 2.} (Color online) One-soliton solutions of the Kundu-NLS equation \eqref{gtc-NLS} with $N=1,\epsilon=0.5, z_{1}=1.5i,A_{+}[z_{1}]=1$:
$\textbf{(a)}$: breather solution with $q_{-}=1$; $\textbf{(b)}$: breather solution with $q_{-}=0.5$; $\textbf{(c)}$: breather solution with $q_{-}=1$;
$\textbf{(d)}$: bright soliton solution with $q_{-}\rightarrow0$.\\}

\noindent
{\rotatebox{0}{\includegraphics[width=4.8cm,height=3.9cm,angle=0]{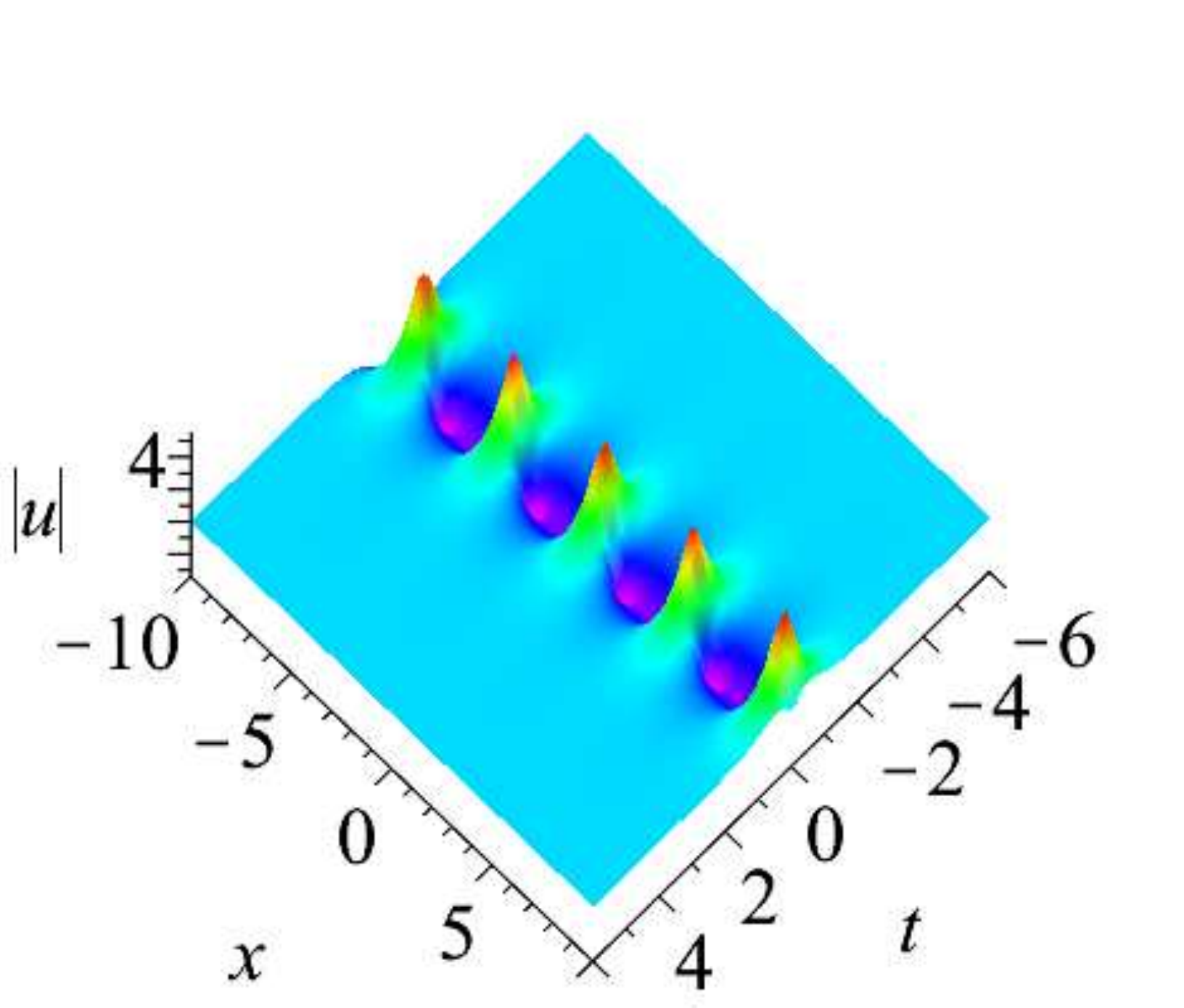}}}
~~~
{\rotatebox{0}{\includegraphics[width=4.8cm,height=3.9cm,angle=0]{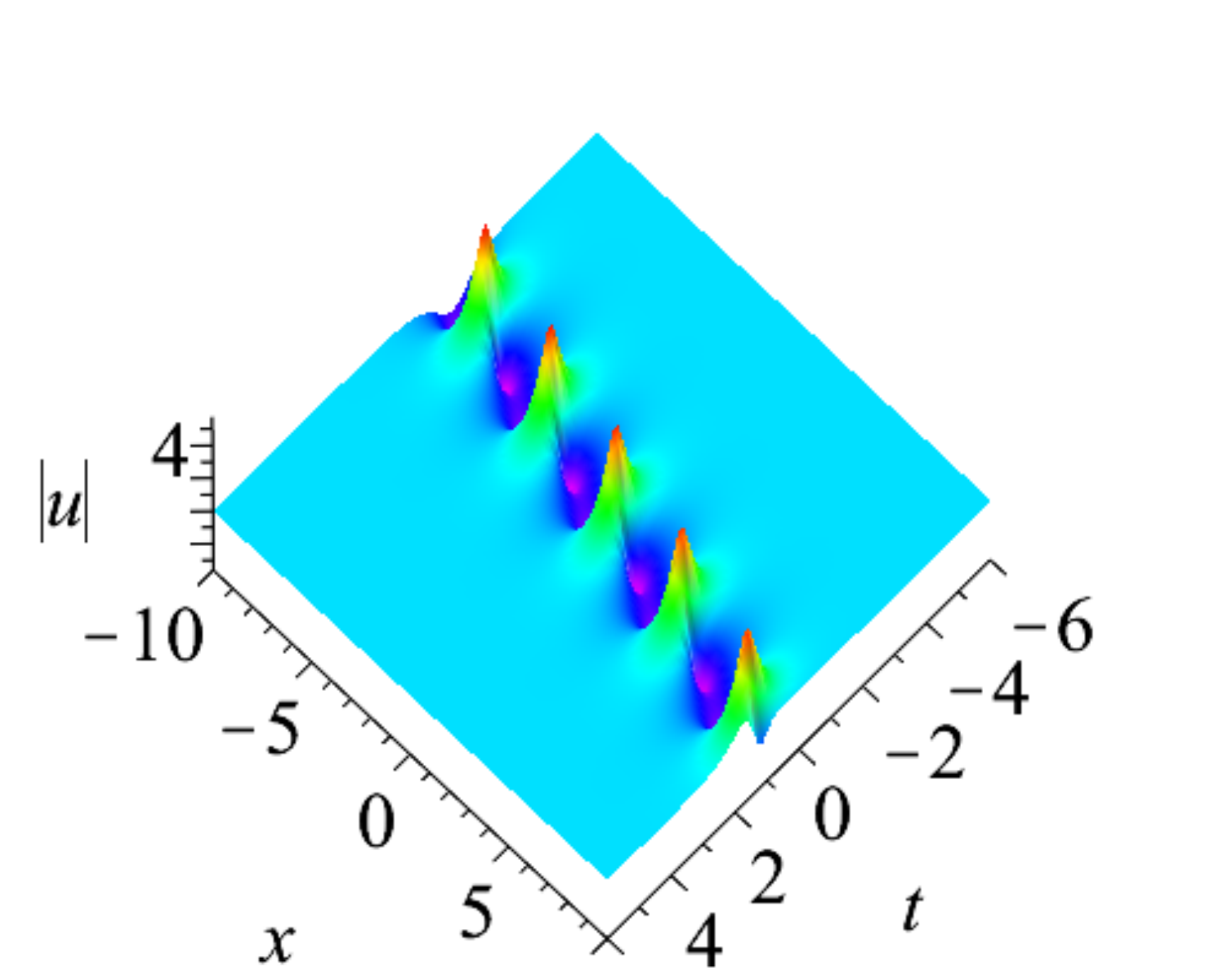}}}
~~~
{\rotatebox{0}{\includegraphics[width=4.8cm,height=3.9cm,angle=0]{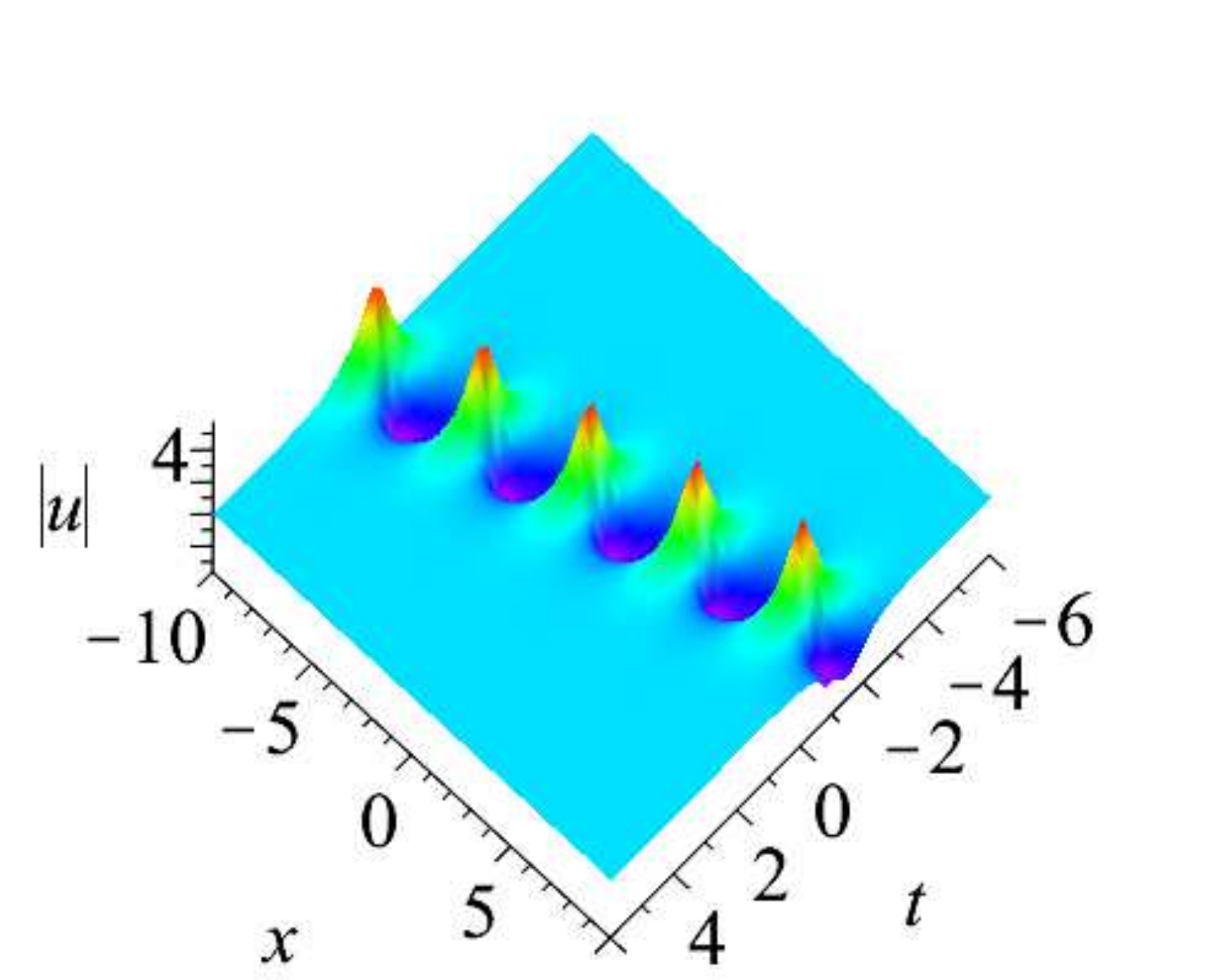}}}

$\qquad\qquad\textbf{(a)}\qquad\qquad\qquad\qquad\qquad\qquad\qquad\textbf{(b)}
\qquad\qquad\qquad\qquad\qquad\qquad\qquad\textbf{(c)}$\\

%\noindent
%{\rotatebox{0}{\includegraphics[width=4.8cm,height=3.9cm,angle=0]{3-4.eps}}}
%~~~
%{\rotatebox{0}{\includegraphics[width=4.8cm,height=3.9cm,angle=0]{3-5.eps}}}
%~~~
%{\rotatebox{0}{\includegraphics[width=4.8cm,height=3.9cm,angle=0]{3-6.eps}}}
%
%$\qquad\qquad\textbf{(d)}\qquad\qquad\qquad\qquad\qquad\qquad\qquad\textbf{(e)}
%\qquad\qquad\qquad\qquad\qquad\qquad\qquad\textbf{(f)}$\\
\noindent { \small \textbf{Figure 3.} (Color online)
One-soliton solutions of the Kundu-NLS equation \eqref{gtc-NLS} with $N=1,A_{+}[z_{1}]=1, \epsilon=0.5$:
$\textbf{(a)}$: $z_{1}=e^{3\pi/4}$; $\textbf{(b)}$: $z_{1}=0.8e^{3\pi/4}$;$\textbf{(c)}$: $z_{1}=-0.8e^{3\pi/4}$.\\}

$~~~~~~$
{\rotatebox{0}{\includegraphics[width=5.2cm,height=3.6cm,angle=0]{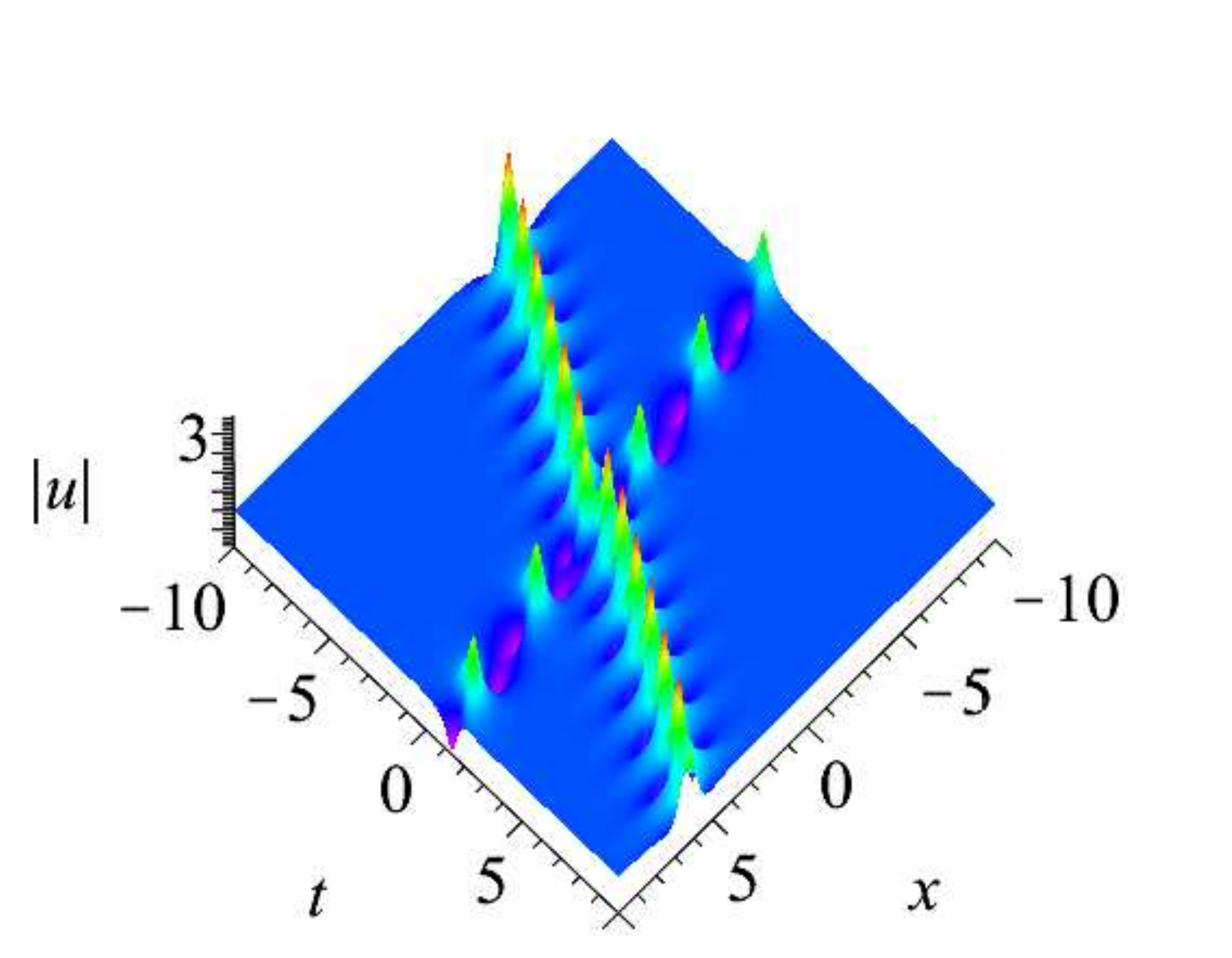}}}
~~~~~~~~~~~~~~~
{\rotatebox{0}{\includegraphics[width=5.2cm,height=3.6cm,angle=0]{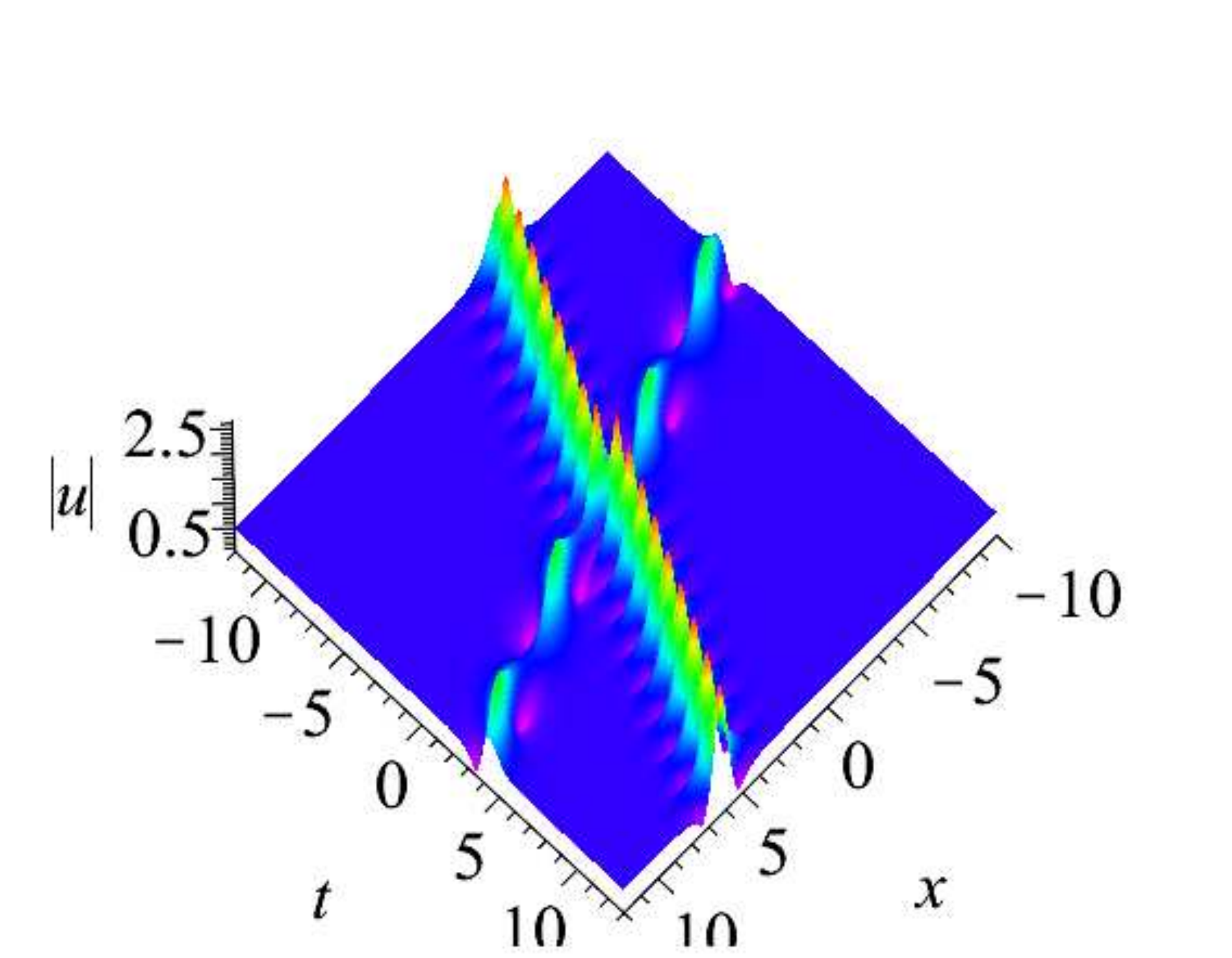}}}

$\qquad\qquad\qquad\qquad\textbf{(a)}
\qquad\qquad\qquad\qquad\qquad\qquad\qquad\qquad\qquad\textbf{(b)}
$\\

$~~~~~~$
{\rotatebox{0}{\includegraphics[width=5.2cm,height=3.6cm,angle=0]{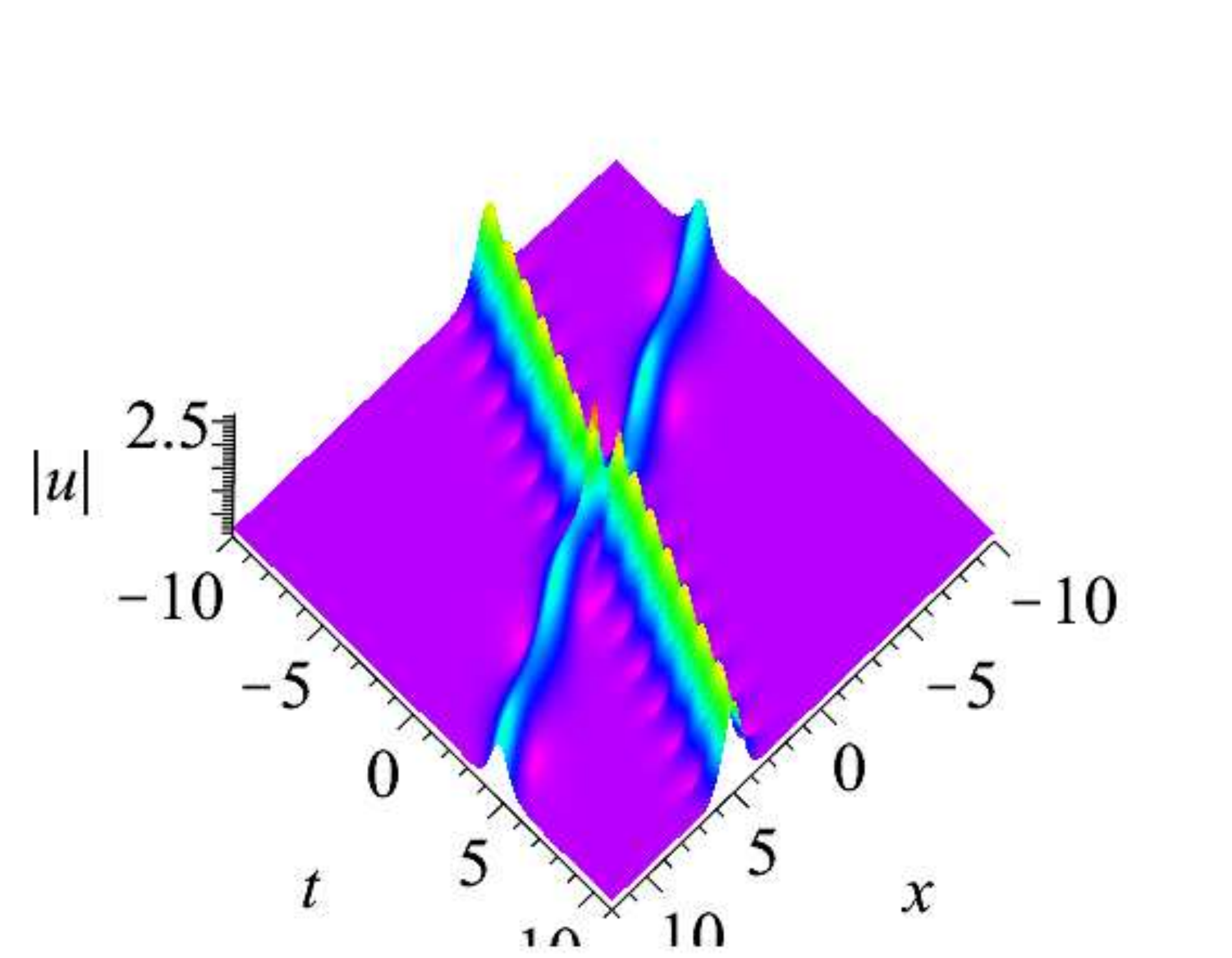}}}
~~~~~~~~~~~~~~~
{\rotatebox{0}{\includegraphics[width=5.2cm,height=3.6cm,angle=0]{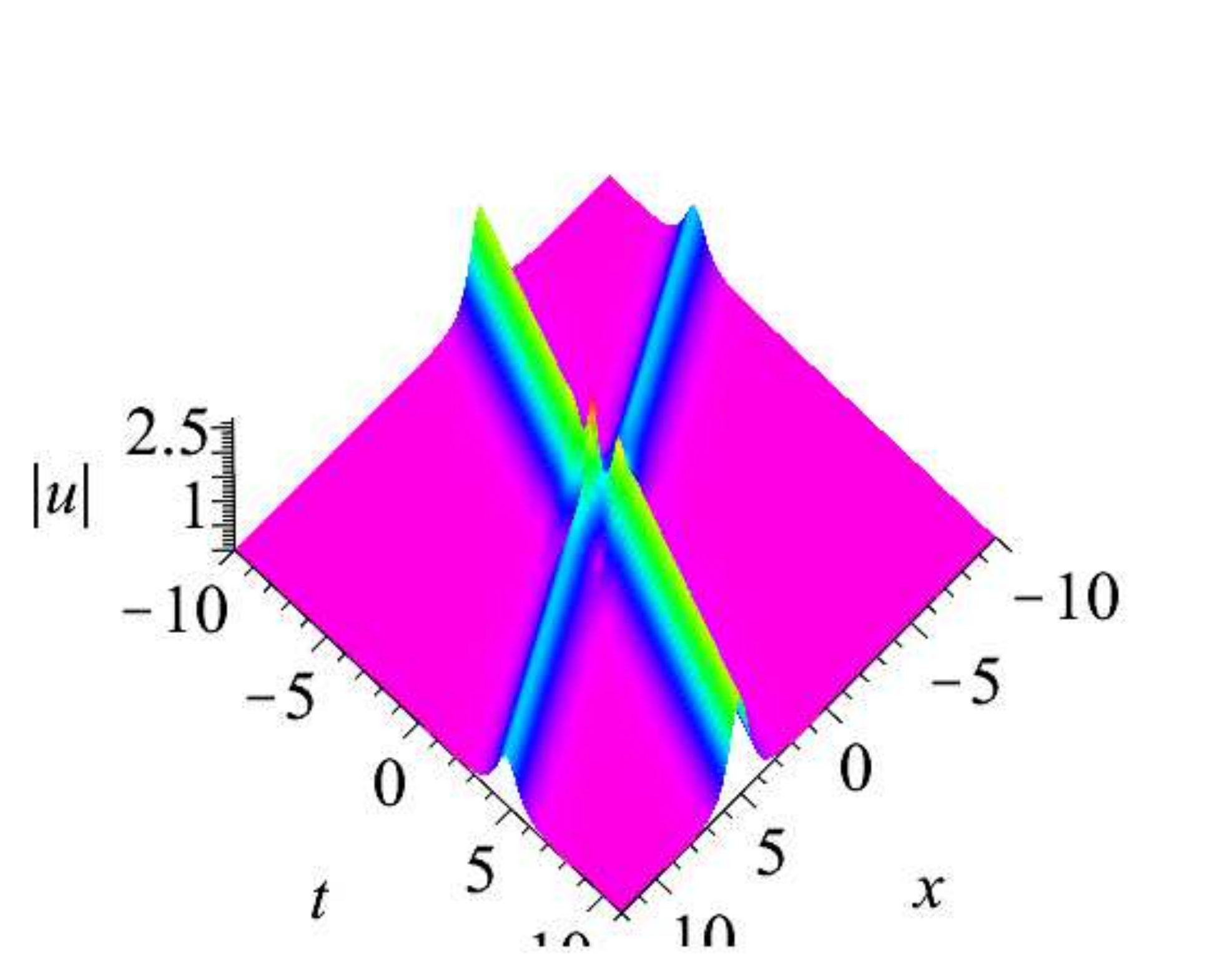}}}

$\qquad\qquad\qquad\qquad\textbf{(c)}
\qquad\qquad\qquad\qquad\qquad\qquad\qquad\qquad\qquad\textbf{(d)}
$\\
\noindent { \small \textbf{Figure 4.} (Color online) Two-soliton solutions of the Kundu-NLS equation \eqref{gtc-NLS} with
$N=2,z_{1}=0.2+2i, \epsilon=0.5, z_{2}=1+i, A_{+}[z_{1}]=1,A_{+}[z_{2}]=1$:
$\textbf{(a)}$:  breather-breather solution with $q_{-}=1$; $\textbf{(b)}$:  breather-breather solution with $q_{-}=0.5$; $\textbf{(c)}$:
 breather-breather solution with $q_{-}=0.2$;
$\textbf{(d)}$: bright-bright soliton solution with $q_{-}\rightarrow0$.\\}

$~~~~$
{\rotatebox{0}{\includegraphics[width=4.2cm,height=3.6cm,angle=0]{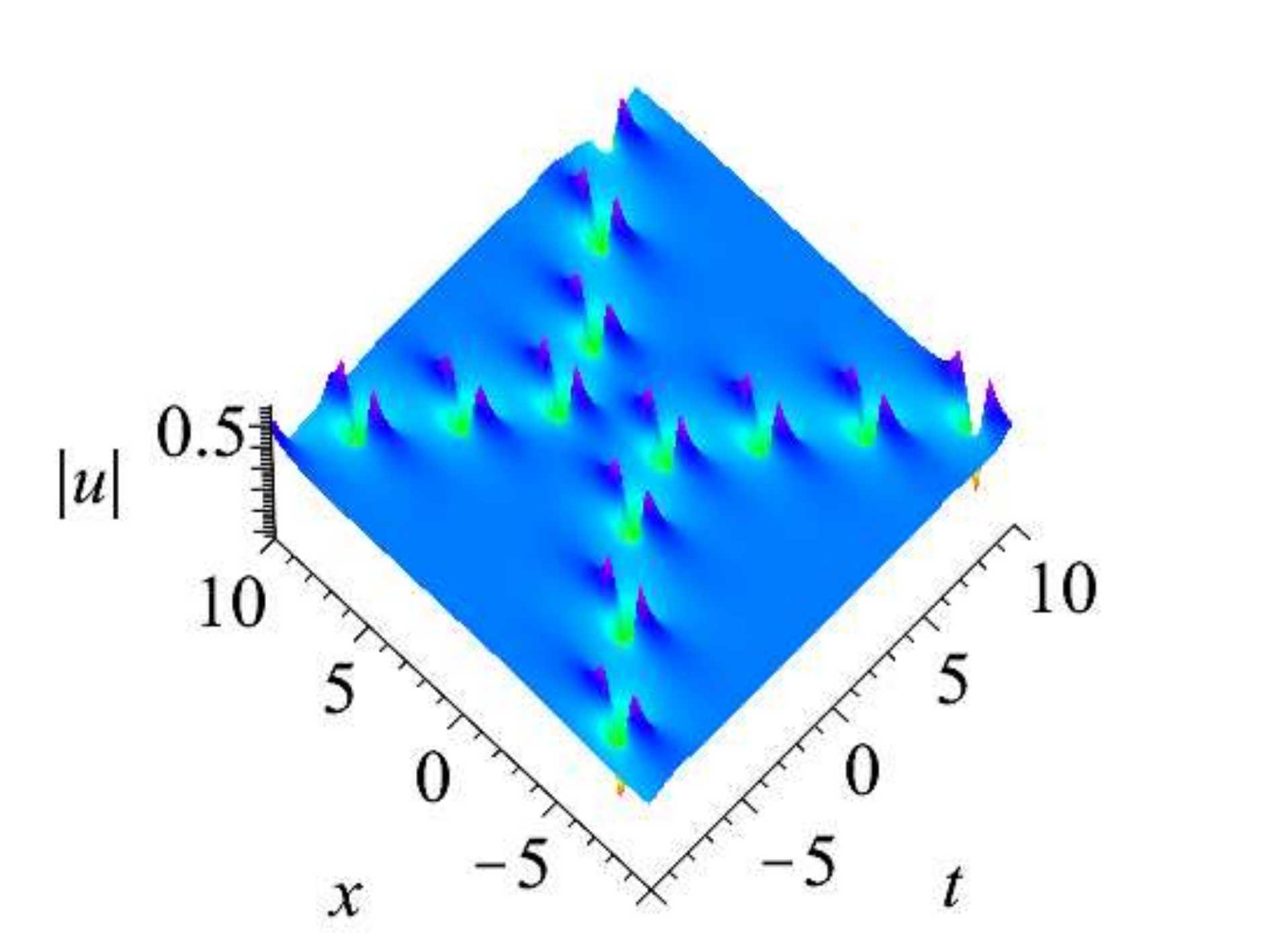}}}
~~~
{\rotatebox{0}{\includegraphics[width=4.2cm,height=3.6cm,angle=0]{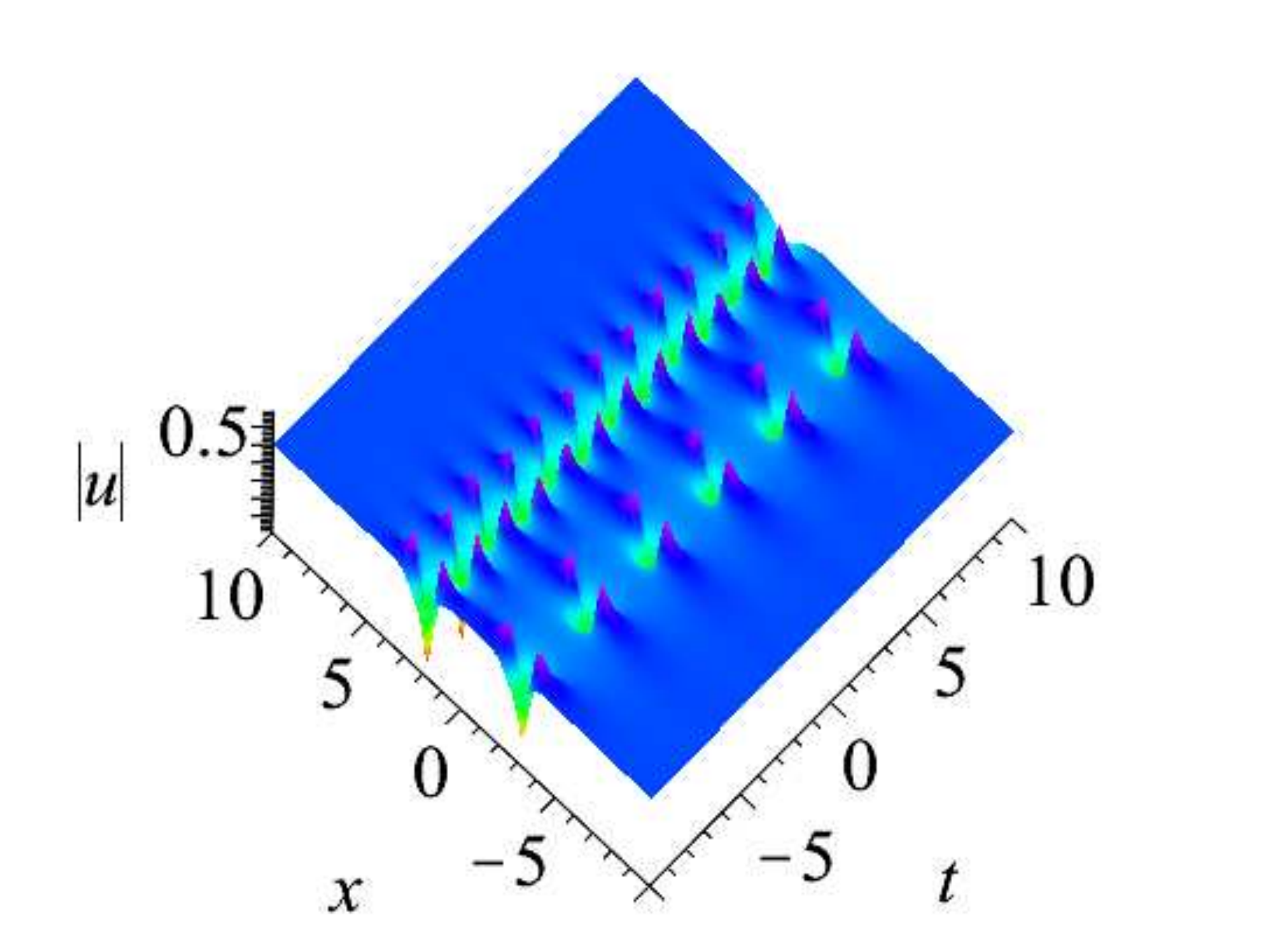}}}
~~~
{\rotatebox{0}{\includegraphics[width=4.2cm,height=3.6cm,angle=0]{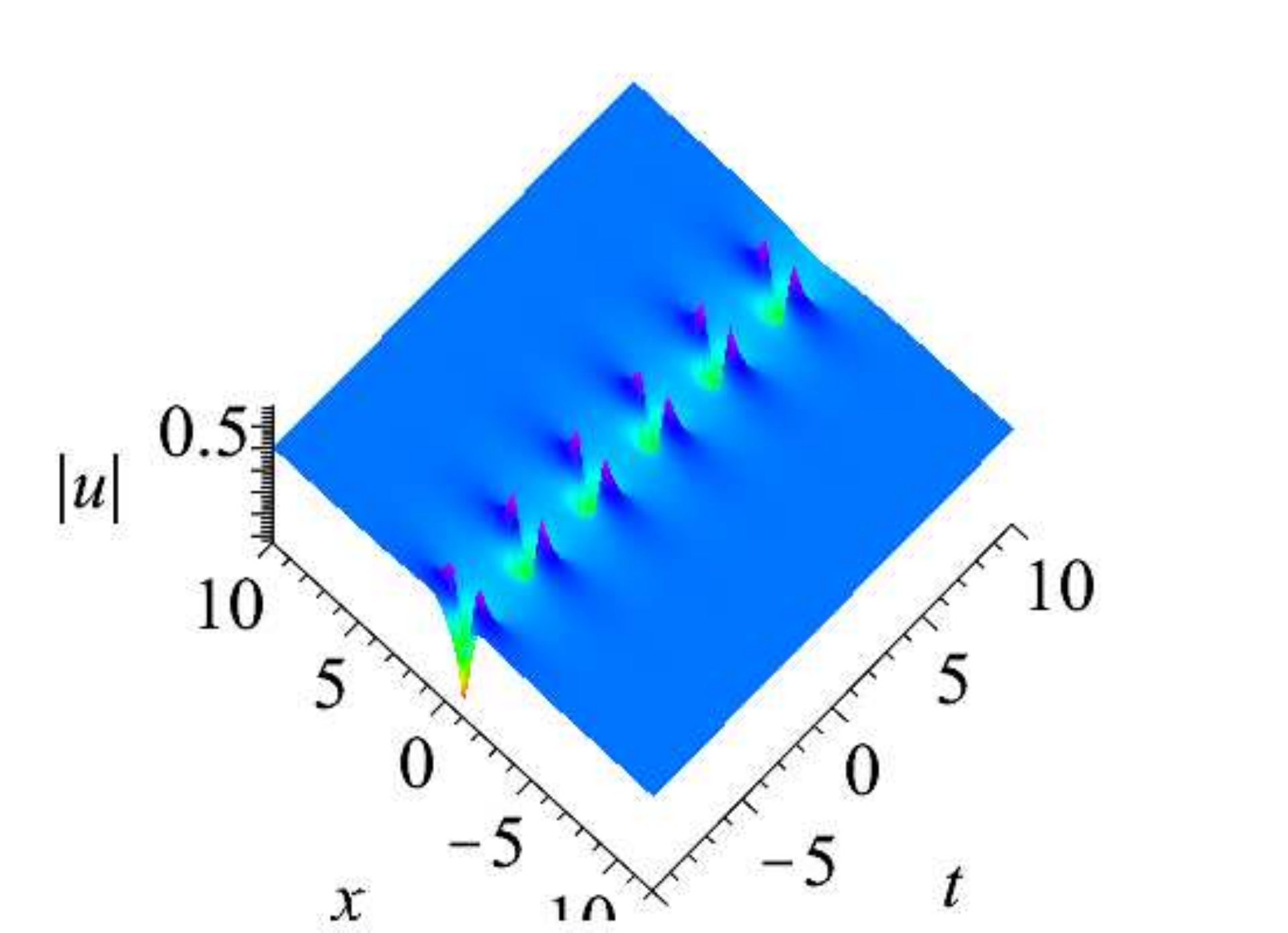}}}

$\qquad\qquad\qquad\textbf{(a)}\qquad\qquad\qquad\qquad\qquad\qquad\quad\textbf{(b)}
\qquad\qquad\qquad\qquad\qquad\qquad\textbf{(c)}$\\
\noindent { \small \textbf{Figure 5.} (Color online)  Two-soliton solutions of the Kundu-NLS equation \eqref{gtc-NLS} with
$N=2, \epsilon=0.5, A_{+}[z_{1}]=1,A_{+}[z_{2}]=1$:$\textbf{(a)}$: $z_{1}=0.2+1.5i,z_{2}=-0.2+1.5i$; $\textbf{(b)}$: $z_{1}=0.5i,z_{2}=1.5i$;
$\textbf{(c)}$: $z_{1}=1.5i,z_{2}=1.5i$.
\\}

$~~~~~~$
{\rotatebox{0}{\includegraphics[width=5.2cm,height=3.6cm,angle=0]{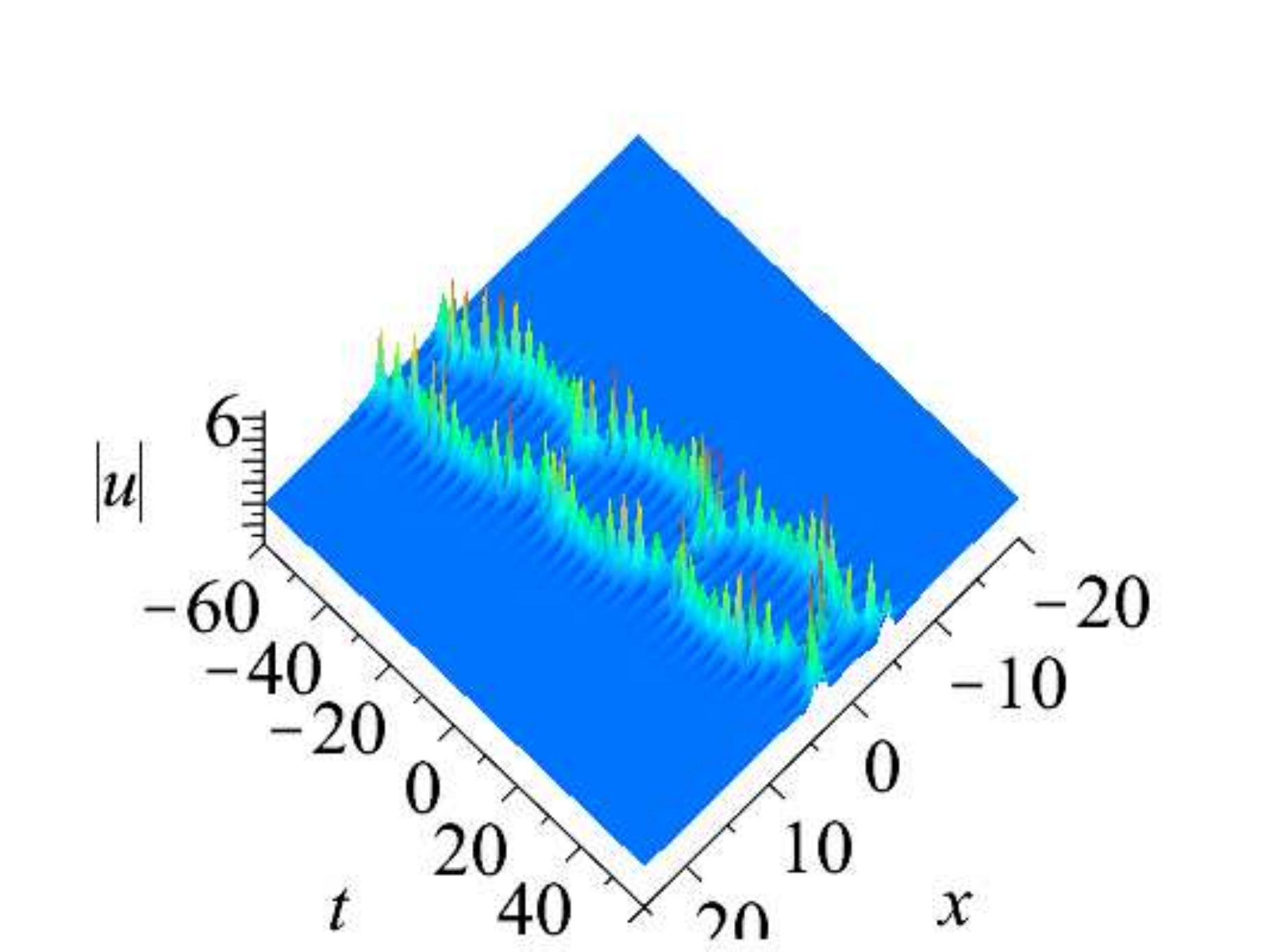}}}
~~~~~~~~~~~~~~~
{\rotatebox{0}{\includegraphics[width=5.2cm,height=3.6cm,angle=0]{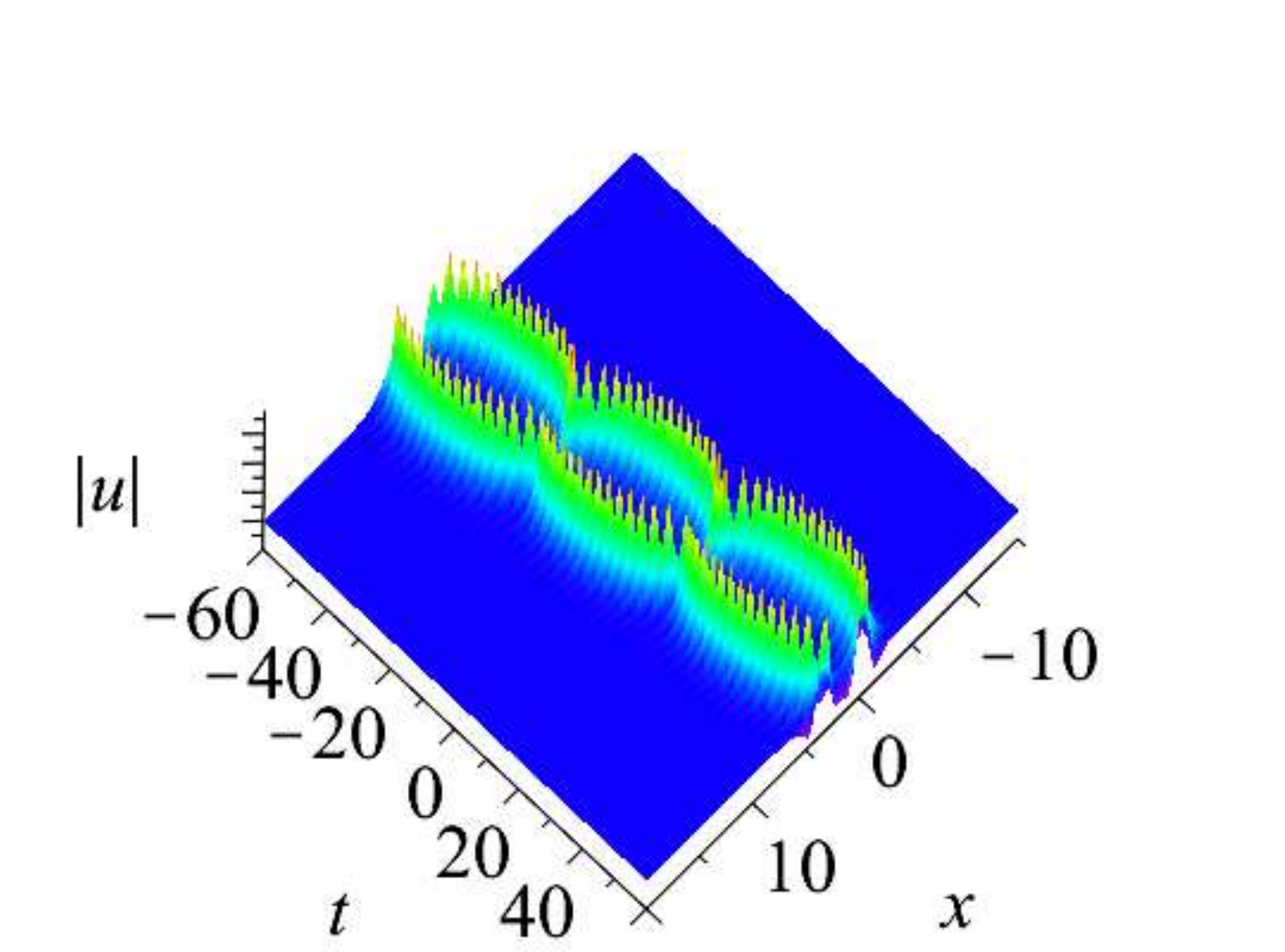}}}

$\qquad\qquad\qquad\qquad\textbf{(a)}
\qquad\qquad\qquad\qquad\qquad\qquad\qquad\qquad\qquad\textbf{(b)}
$\\

$~~~~~~$
{\rotatebox{0}{\includegraphics[width=5.2cm,height=3.6cm,angle=0]{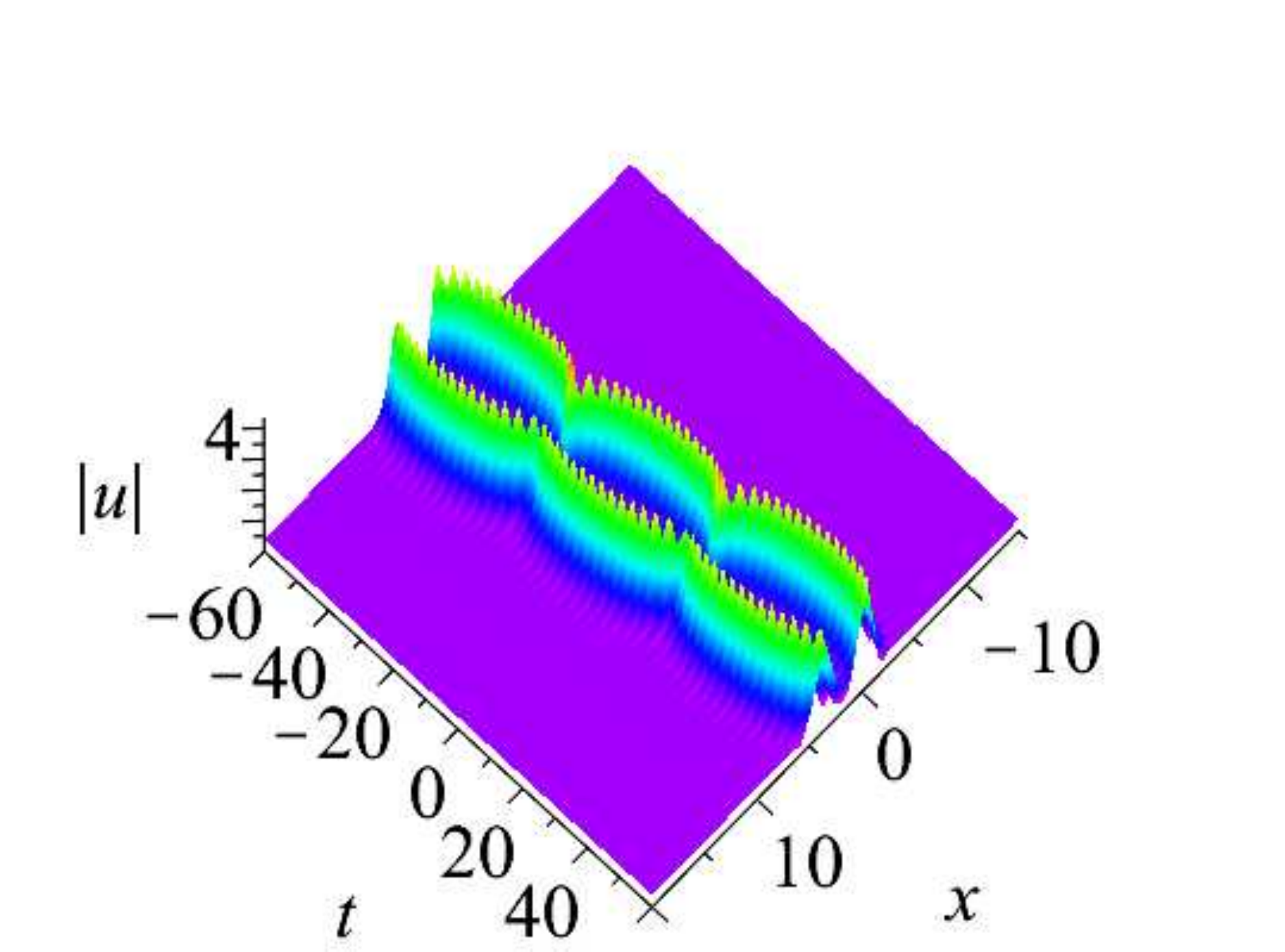}}}
~~~~~~~~~~~~~~~
{\rotatebox{0}{\includegraphics[width=5.2cm,height=3.6cm,angle=0]{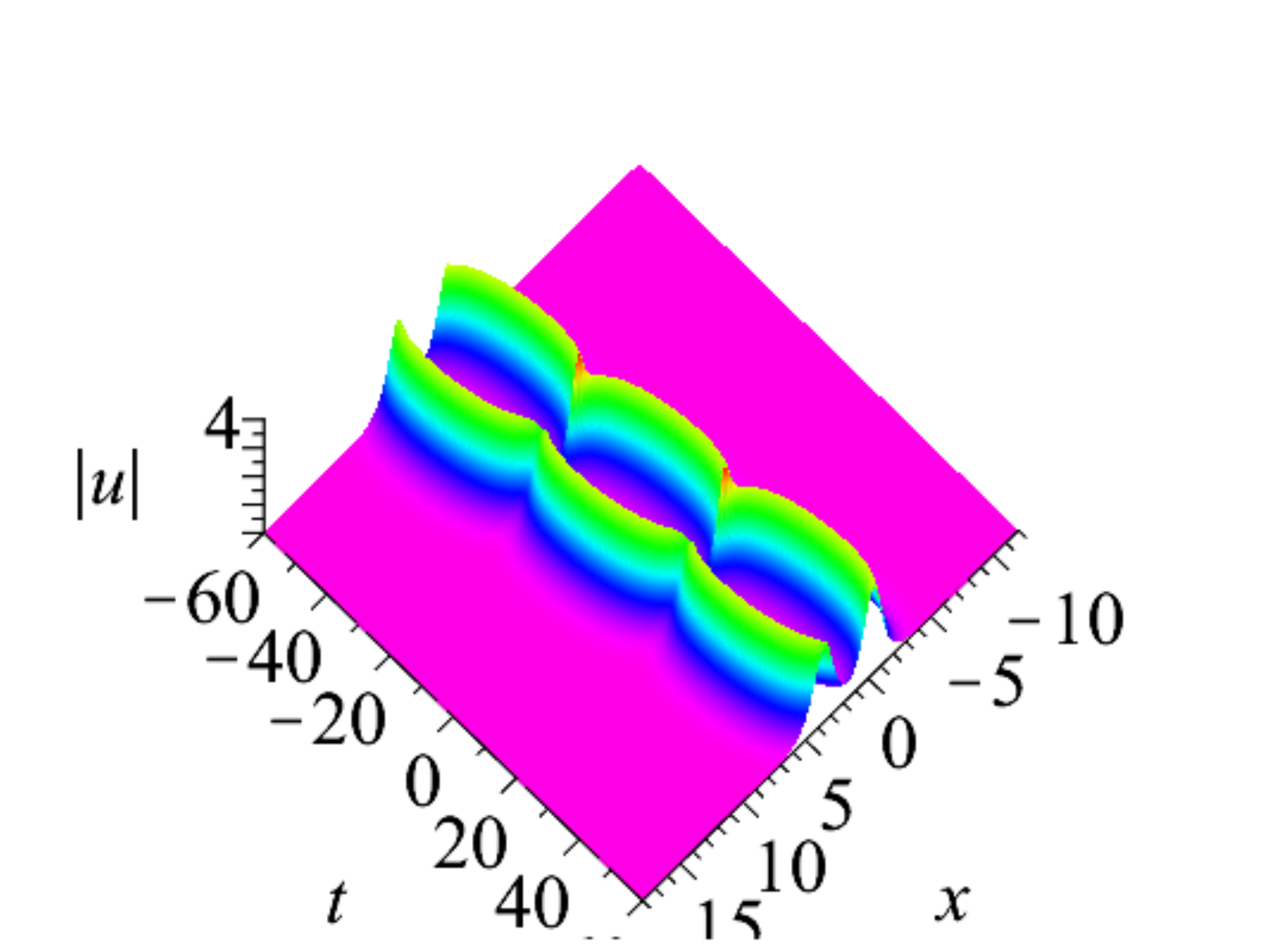}}}

$\qquad\qquad\qquad\qquad\textbf{(c)}
\qquad\qquad\qquad\qquad\qquad\qquad\qquad\qquad\qquad\textbf{(d)}
$\\
\noindent { \small \textbf{Figure 6.} (Color online) Two-soliton solutions of the Kundu-NLS equation \eqref{gtc-NLS} with
$N=2,z_{1}=-1/65+1.5i, z_{2}=-1/65+1.05i, A_{+}[z_{1}]=1,A_{+}[z_{2}]=1,\epsilon=0.5$:
$\textbf{(a)}$: breather-breather solution with $q_{-}=1$; $\textbf{(b)}$: breather-breather solution with $q_{-}=0.5$;
$\textbf{(c)}$: breather-breather solution with $q_{-}=1$;
$\textbf{(d)}$: bright-bright soliton solution with $q_{-}\rightarrow0$.\\}

\section{The Kundu-NLS equation with NZBCs: Double poles}
In the section, we assume that the discrete spectral points $Z^{f}$ are double zeros of the scattering coefficients
$s_{11}(z)$ and $s_{22}(z)$, i.e., $s_{11}(z_{0})=s'_{11}(z_{0})\neq 0$ for $\forall z_{0}\in Z^{f}\bigcap D_{+}^{f}$,
and $s_{22}(z_{0})=s'_{22}(z_{0})=0$, $s''_{22}(z_{0})\neq0$ for $\forall z_{0}\in Z^{f}\bigcap D_{-}^{f}$.
Convenience sake, we present a simple proposition \cite{MP-2014}: if $f(z)$ and $g(z)$ are analytic in some complex domain $\Omega$,
and $z_{0}\in\Omega$ represents a double zero of $g(z)$ and $f(z_{0})\neq0$,
then the function $f(z)/g(z)$ admits the double pole $z=z_{0}$,
and its residue $\mbox{Res}[f/g]$ and the coefficient $P_{-2}[f/g]$ of $(z-z_{0})^{-2}$ in the Laurent series are expressed by
\begin{align}\label{DP-1}
\mathop{P_{-2}}\limits_{z=z_{0}}\left[\frac{f}{g}\right]=2f(z_{0})/g''(z_{0}),
~~\mathop{\mbox{Res}}\limits_{z=z_{0}}\left[\frac{f}{g}\right]
=2\left(\frac{f'(z_{0})}{g''(z_{0})}-\frac{f(z_{0}g'''(z_{0}))}{3[g''(z_{0}]^2}\right).
\end{align}
Similar to \cite{MP-2014}, for $s_{11}(z_{0})=s'_{11}(z_{0})=0$, $s'_{11}(z_{0})\neq0$ in $\forall z_{0}\in Z^{f}\bigcap D_{+}^{f}$.
Eq.\eqref{ISP-6} still holds. The first expression of \eqref{Lax-15} can be rewritten as
\begin{equation}\label{DP-2}
s_{11}(z)\gamma_{f}(z)=\left|\phi_{+1}(x,t,z),\phi_{-2}(x,t,z)\right|,
\end{equation}
whose first-order partial derivative with respect to $z$ reads
\begin{equation}\label{DP-3}
[s_{11}(z)\gamma_{f}(z)]'=\left|\phi'_{+1}(x,t,z),\phi_{-2}(x,t,z)\right|+\left|\phi_{+1}(x,t,z),\phi'_{-2}(x,t,z)\right|.
\end{equation}
Choosing $z=z_{0}\in Z^{f}\bigcap D_{+}^{f}$ in \eqref{DP-3} and utilizing $s_{11}(z_{0})=s'_{11}(z_{0})=0$ and \eqref{ISP-8} can lead to
\begin{equation}\label{DP-4}
\left|\phi'_{+1}(x,t,z_{0})-b_{+}(z_{0})\phi'_{-2}(x,t,z_{0}),\phi_{-2}(x,t,z_{0})\right|=0,
\end{equation}
which means that there exists another constant $c_{+}(z_{0})$ such that
\begin{equation}\label{DP-5}
\phi'_{+1}(x,t,z_{0})=c_{+}(z_{0})\phi_{-2}(x,t,z_{0})+b_{+}(z_{0})\phi'_{-2}(x,t,z_{0}).
\end{equation}
It follows from \eqref{ISP-8}, \eqref{DP-1} and \eqref{DP-5} that
\begin{align}\label{DP-6}
&\mathop{P_{-2}}\limits_{z=z_{0}}\left[\frac{\phi_{+}(x,t,z)}{s_{11}(z)}\right]
=\frac{2\phi_{+1}(x,t,z_{0})}{s''_{11}(z_{0})}=\frac{2b_{+}(z_{0})}{s''_{11}(z_{0})}\phi_{-2}(x,t,z_{0})=A_{+}\left[z_{0}\right]\phi_{-2}(x,t,z_{0}),
\notag\\
&\mathop{\mbox{Res}}\limits_{z=z_{0}}\left[\frac{\phi_{+1}(x,t,z)}{s_{11}(z)}\right]=\frac{2\phi'_{+1}(x,t,z_{0})}{s''_{11}(z_{0})}
-\frac{2\phi_{+1}(x,t,z_{0})s'''_{11}(z_{0})}{3(s''_{11}(z_{0}))^{2}}\notag\\
&~~~~~~~~=A_{+}\left[z_{0}\right]\left[\phi'_{-2}(x,t,z_{0})+B_{+}[z_{0}]\phi_{-2}(x,t,z_{0})\right].
\end{align}

Following a similar way , for $s_{22}(z^{*}_{0})=s'_{22}(z^{*}_{0})=0$, $s''_{22}(z^{*}_{0})\neq0$ in $\forall z^{*}_{0}\in Z^{f}\bigcap D_{-}^{f}$,
Eq.\eqref{ISP-8} holds.
According to the second one of \eqref{Lax-15} and \eqref{ISP-8}, we have
\begin{equation}\label{DP-7}
\phi'_{+2}(x,t,z^{*}_{0})=c_{-}(z^{*}_{0})\phi_{-1}(x,t,z^{*}_{0})+b_{-}(z^{*}_{0})\phi'_{-1}(x,t,z^{*}_{0})
\end{equation}
for $c_{-}(z^{*}_{0})$.

It follows from \eqref{ISP-8}, \eqref{DP-1} and \eqref{DP-7} that
\begin{equation}\label{DP-8}
\left\{ \begin{aligned}
&\mathop{P_{-2}}\limits_{z=z^{*}_{0}}\left[\frac{\phi_{+2}(x,t,z)}{s_{22}(z)}\right]=\frac{2\phi_{+2}(x,t,z^{*}_{0})}{s''_{22}(z^{*}_{0})}
=\frac{2b_{-}(z^{*}_{0})}{s''_{22}(z^{*}_{0})}\phi_{-1}(x,t,z^{*}_{0})=A_{-}[z^{*}_{0}]\phi_{-1}(x,t,z^{*}_{0}),\\
&\mathop{\mbox{Res}}\limits_{z=z^{*}_{0}}\left[\frac{\phi_{+2}(x,t,z)}{s_{22}(z)}\right]=A_{-}[z^{*}_{0}]\left[\phi'_{-1}(x,t,z^{*}_{0})
+B_{-}[z^{*}_{0}]\phi_{-1}(x,t,z^{*}_{0})\right].
           \end{aligned} \right.
\end{equation}
We thus know
\begin{equation}\label{DP-9}
\left\{ \begin{aligned}
&A_{+}[z_{0}]=\frac{2b_{+}[z_{0}]}{s''_{11}(z_{0})},~~B_{+}[z_{0}]=\frac{c_{+}[z_{0}]}{b_{+}[z_{0}]}-\frac{s'''_{11}(z_{0})}{3s''_{11}(z_{0})},~~
z_{0}\in Z^{f}\cap D_{+}^{f},\\
&A_{+}[z^{*}_{0}]=\frac{2b_{+}[z^{*}_{0}]}{s''_{11}(z^{*}_{0})},~~B_{+}[z^{*}_{0}]=\frac{c_{+}[z^{*}_{0}]}
{b_{+}[z^{*}_{0}]}-\frac{s'''_{11}(z^{*}_{0})}{3s''_{11}(z^{*}_{0})},~~
z^{*}_{0}\in Z^{f}\cap D_{-}^{f},
           \end{aligned} \right.
\end{equation}
from which we obtain the following symmetries
\begin{align}\label{DP-10}
&A_{+}[z_{0}]=-A^{*}_{-}[z^{*}_{0}]=\frac{z_{0}^4q^{*}_{-}}{q_{0}^4q_{-}}A_{-}\left[-\frac{q_{0}^2}{z_{0}}\right],\notag\\
&B_{+}[z_{0}]=B^{*}_{-}[z^{*}_{0}]=\frac{q_{0}^2}{z_{0}^2}B_{-}\left[-\frac{q_{0}^2}{z_{0}}\right]+\frac{2}{z_{0}},~~z_{0}\in Z^{f}\cap D_{+}^{f},
\end{align}
which arrive at
\begin{equation}\label{DP-11}
\left\{ \begin{aligned}
&A_{+}[z_{n}]=-A^{*}_{-}[z^{*}_{n}]=-\frac{z_{n}^{4}q^{*}_{-}}{q_{0}^4q_{-}}A_{+}\left[-\frac{q_{0}^2}{z^{*}_{n}}\right],~~~~z_{n}\in Z^{f}\cap D_{+}^{f},\\
&B_{+}[z_{n}]=B^{*}_{-}[z^{*}_{n}]=\frac{q_{0}^2}{z_{n}^2}B^{*}_{+}\left[-\frac{q_{0}^2}{z^{*}_{n}}\right]+\frac{2}{z_{n}},~~~~~z^{*}_{n}\in Z^{f}\cap D_{-}^{f}.
           \end{aligned} \right.
\end{equation}
To sum up, we have
\begin{equation}\label{DP-12}
\left\{ \begin{aligned}
&\mathop{P_{-2}}\limits_{z=\xi_{n}} M_{1}^{+}(x,t,z)=P_{-2}\left[\frac{\mu_{+1}(x,t,z)}{s_{11}(z)}\right]=A[\xi_{n}]e^{-2i\theta(x,t,\xi_{n})}\mu_{-2}(x,t,\xi_{n}),\\
&\mathop{P_{-2}}\limits_{z=\widehat{\xi}_{n}}M_{2}^{-}(x,t,z)=P_{-2}\left[\frac{\mu_{+2}(x,t,z)}{s_{22}(z)}\right]=A\left[\widehat{\xi}_{n}\right]
e^{2i\theta(x,t,\widehat{\xi}_{n})}\mu_{-1}(x,t,\xi_{n}),\\
&\mathop{\mbox{Res}}\limits_{z=\xi_{n}}M_{1}^{+}(x,t,z)=\mbox{Res}\left[\frac{\mu_{+1}(x,t,z)}{s_{11}(z)}\right]\\
&~~~~~~~~~~~~~~~~~~~~~~~~=A\left[\xi_{n}\right]e^{-2i\theta(x,t,\xi_{n})}\left\{\mu'_{-2}(x,t,\xi_{n})
+\left[B\left[\xi_{n}\right]-2i\theta'(x,t,\xi_{n})\right]\mu_{-2}(x,t,\xi_{n})\right\},\\
&\mathop{\mbox{Res}}\limits_{z=\widehat{\xi}_{n}}M_{2}^{-}(x,t,z)=\mbox{Res}\left[\frac{\mu_{+2}(x,t,z)}{s_{22}(z)}\right]\\
&~~~~~~~~~~~~~~~~~~~~~~~~=A[\widehat{\xi}_{n}]e^{2i\theta(x,t,\widehat{\xi}_{n})}\left\{\mu'_{-1}(x,t,\widehat{\xi}_{n})+\left[B\left[\widehat{\xi}_{n}\right]
+2i\theta'(x,t,\widehat{\xi}_{n})\right]\mu_{-1}(x,t,\widehat{\xi}_{n})\right\}.
           \end{aligned} \right.
\end{equation}
The RHP in Proposition 3.1 still holds for the case of double poles. To solve this kind of RHP, we must subtract out the asymptotic values as $z\rightarrow\infty$ and $z\rightarrow0$ and the singularity contributions
\begin{align}\label{DP-13}
&M_{dp}(x,t,z)=I+\frac{i}{z}\sigma_{3}Q_{-}+\sum_{n=1}^{2N}M_{dp}^{n},\notag\\
&M_{dp}^{n}=\frac{\mathop{P_{-2}}\limits_{z=\xi_{n}}M^{+}}{(z-\xi_{n})^2}+\frac{\mathop{P_{-2}}\limits_{z=\widehat{\xi}_{n}}M^{-}}{(z-\widehat{\xi}_{n})^2}
+\frac{\mathop{\mbox{Res}}\limits_{z=\xi_{n}} M^{+}}{z-\xi_{n}}+\frac{\mathop{\mbox{Res}}\limits_{z=\widehat{\xi}_{n}} M^{+}}{z-\widehat{\xi}_{n}}.
\end{align}
Then it follows from the jump condition $M^{-}=M^{+}(I-J)$ that
\begin{equation}\label{DP-14}
M^{-}(x,t,z)-M_{dp}(x,t,z)=M^{+}(x,t,z)-M_{dp}(x,t,z)-M^{+}(x,t,z)J,
\end{equation}
where $M^{\pm}(x,t,z)-M_{dp}(x,t,z)$ are analytic in $D_{\pm}^{f}$.
Additionally, their asymptotics are both $O(1/z)$ as $z\rightarrow\infty$ and $O(1)$ as $z\rightarrow0$,
and $J(x,t,z)$ is $O(1/z)$ as $z\rightarrow\infty$, and $O(z)$ as $z\rightarrow0$.
Consequently, the Cauchy projectors and Plemelj's formulae are used to solve
\eqref{DP-14} to generate
\begin{equation}\label{DP-15}
M(x,t,z)=M_{dp}(x,t,z)+\frac{1}{2\pi i}\int_{\Sigma^{f}}\frac{M^{+}(x,t,\zeta)J(x,t,\zeta)}{\zeta-z}d\zeta,~~z\in\mathbb{C}\backslash\Sigma^{f},
\end{equation}
where $\int_{\Sigma^{f}}$ represents the integral along the oriented contours displayed in Fig.1 (right).

Thus, by using \eqref{DP-12}, the parts of $P_{-2}(\cdot)$  and $\mbox{Res}(\cdot)$ in \eqref{DP-15} can be expressed as
\begin{align}\label{DP-16}
M_{dp}^{n}=
\left(C_{n}(z)\left[\mu'_{-2}(\xi_{n})+\left(D_{n}+\frac{1}{z-\xi_{n}}\right)\mu_{-2}(\xi_{n})\right],
\widehat{C}_{n}(z)\left[\mu'_{-1}\left(\widehat{\xi}_{n}\right)+\left(D_{n}+\frac{1}{z-\widehat{\xi}_{n}}\right)\mu_{-2}(\widehat{\xi}_{n})\right]\right),
\end{align}
where
\begin{align}\label{DP-17}
&C_{n}(z)=\frac{A_{+}[\xi_{n}]}{z-\xi_{n}}e^{-2i\theta(\xi_{n})},~~D_{n}=B_{+}[\xi_{n}]-2i\theta'(\xi_{n}),\\
&\widehat{C}_{n}(z)=\frac{A_{-}[\widehat{\xi}_{n}]}{z-\widehat{\xi}_{n}}e^{2i\theta(\widehat{\xi}_{n})},
~~\widehat{D}_{n}=B_{-}(\widehat{\xi}_{n})+2i\theta'(\widehat{\xi}_{n}).
\end{align}
To obtain $\mu'_{-2}(\xi_{n})$, $\mu_{-2}(\xi_{n})$, $\mu'_{-1}$, and $\mu_{-1}(\xi_{n})$ in \eqref{DP-16},
for $z=\xi_{s}(s=1,2,\ldots,2N)$,
According the second column of $M(x,t,\lambda)$ expressed by \eqref{DP-15} and \eqref{DP-16}, we have
\begin{equation}\label{DP-18}
\mu_{-2}(z)=\left(
              \begin{array}{c}
                \frac{iq_{-}}{z} \\
                1 \\
              \end{array}
            \right)+\sum_{n=1}^{2N}\widehat{C}_{n}(z)\left[\mu'_{-1}(\widehat{\xi}_{n})
            +\left(\widehat{D}_{n}+\frac{1}{z-\widehat{\xi}_{n}}\right)\mu_{-1}(\widehat{\xi}_{n})\right]+\frac{1}{2\pi i}\int_{\Sigma^{f}}\frac{(M^{+}J)_{2}(\zeta)}{\zeta-z}d\zeta,
\end{equation}
whose first-order derivative with respect to $z$ can lead to
\begin{equation}\label{DP-19}
\mu'_{-2}(z)=\left(
              \begin{array}{c}
                -\frac{iq_{-}}{z^2} \\
                1 \\
              \end{array}
            \right)-\sum_{n=1}^{2N}\frac{\widehat{C}_{n}(z)}{z-\widehat{\xi}_{n}}\left[\mu'_{-1}(\widehat{\xi}_{n})
            +\left(\widehat{D}_{n}+\frac{1}{z-\widehat{\xi}_{n}}\right)\mu_{-1}(\widehat{\xi}_{n})\right]+\frac{1}{2\pi i}\int_{\Sigma^{f}}\frac{(M^{+}J)_{2}(\zeta)}{(\zeta-z)^2}d\zeta.
\end{equation}
Additionally, in view of \eqref{ISP-1}, we obtain
\begin{equation}\label{DP-20}
\mu'_{-2}(z)=-\frac{iq_{-}}{z^2}\mu_{-}\left(\frac{-q_{0}^2}{z}\right)+\frac{iq_{0}^2}{z}\frac{q_{-}}{z^2}\mu'_{-1}\left(-\frac{q_{0}^2}{z}\right).
\end{equation}
Then plugging \eqref{DP-20} into \eqref{DP-18} and \eqref{DP-19} yields
\begin{align}\label{DP-21}
&\sum_{n=1}^{2N}\widehat{C}_{n}(\xi_{s})\mu'_{-1}(\widehat{\xi}_{n})+\left[\widehat{C}_{n}(\xi_{k})\left(\widehat{D}_{n}
+\frac{1}{\xi_{s}-\widehat{\xi}_{n}}\right)-\frac{iq_{-}}{\xi_{s}}\delta_{sn}\right]\mu_{-1}(\widehat{\xi}_{n})\notag\\
&~~~~~~~~~~~~~~=-\left(
    \begin{array}{c}
      \frac{iq_{-}}{\xi_{s}} \\
      1 \\
    \end{array}
  \right)-\frac{1}{2\pi i}\int_{\Sigma^{f}}\frac{(M^{+}J)_{2}(\zeta)}{\zeta-\xi_{k}}d\zeta,\notag\\
&\sum_{n=1}^{2N}\left(\frac{\widehat{C}_{n}(\xi_{s})}{\xi_{s}-\widehat{\xi}_{n}}+\frac{iq_{0}^2q_{-}}{\xi_{s}^3\delta_{sn}}\right)
\mu'_{-1}(\widehat{\xi}_{n})
+\left[\frac{\widehat{C}_{n}(\xi_{s})}{\xi_{s}-\widehat{\xi}_{n}}\left(\widehat{D}_{n}+\frac{2}{\xi_{s}-\widehat{\xi}_{n}}\right)
-\frac{iq_{-}}{\xi_{s}^2}\delta_{sn}\right]
\mu_{-1}(\widehat{\xi}_{n})\notag\\&~~~~~~~~~~~~~~=\left(
                               \begin{array}{c}
                                 -\frac{q_{-}}{\xi_{s}^2} \\
                                 0 \\
                               \end{array}
                             \right)+\int_{\Sigma^{f}}\frac{(M^{+}J)_{2}(\zeta)}{2\pi i(\zeta-\xi_{k})^2}d\zeta,
\end{align}
and we can obtain $\mu_{-}(x,t,\widehat{\xi}_{n})$, $\mu'_{-1}(x,t,\widehat{\xi}_{n}), n=1,2,\ldots,2N$
such that one can also find $\mu_{-2}(x,t,\xi_{n})$, $\mu'_{-2}(x,t,\xi_{n}), n=1,2,\ldots,2N$ form \eqref{DP-20}
As a result, substituting them into \eqref{DP-16}, and then substituting \eqref{DP-16} into \eqref{DP-15}
can lead to the $M(x,t,z)$ in view of the scattering data.

From \eqref{DP-15} and \eqref{DP-16}, we find that
that the asymptotic behavior of $M(x,t,z)$ is still of the form \eqref{JSP-27}.
However, $M^{(1)}(x,t)$ is replaced by
\begin{align}\label{DP-22}
M^{(1)}(x,t)&=i\sigma_{3}Q_{-}-\frac{1}{2\pi i}\int_{\Sigma^{f}}(M^{+}J)(\zeta)d\zeta\notag\\
&+\sum_{n=1}^{2N}\left[A_{+}[\xi_{n}]e^{-2i\theta(\xi_{n})}\left(\mu'_{-2}(\xi_{n})+D_{n}\mu_{-2}(\xi_{n})\right),
A_{-}\left[\widehat{\xi}_{n}\right]e^{2i\theta(\widehat{\xi}_{n})}\left(\mu'_{-2}(\widehat{\xi}_{n})
+\widehat{D}_{n}\mu_{-1}(\widehat{\xi}_{n})\right)\right].
\end{align}

Summarizing the above results, the following property for the potential $u(x,t)$ for the case of double poles holds.\\

\noindent
\textbf{Proposition 4.1.} The potential with double poles of the Kundu-NLS equation \eqref{gtc-NLS} with NZBCs is expressed by
\begin{equation}\label{DP-16}
\epsilon ue^{i\gamma}=q_{-}-i\sum_{n=1}^{2N}A_{-}[\widehat{\xi}_{n}]e^{2i\theta(\xi_{n})}
\left(\mu'_{-11}(\widehat{\xi}_{n})+\widehat{D}_{n}\mu_{-11}(\widehat{\xi}_{n})\right)
+\frac{1}{2\pi}\int_{\Sigma'}\left(M^{+}J\right)_{12}(\zeta)d\zeta,
\end{equation}
where $\widehat{C}_{n}(z)=\frac{A[\widehat{\xi}_{n}]}{z-\xi_{n}}e^{2i\theta(\widehat{\xi}_{n})}$,
$\widehat{D}_{n}=B[\widehat{\xi}_{n}]+2i\theta'(\xi_{n})$,
and $\mu_{-11}(\widehat{\xi}_{n})$ and $\mu'_{-11}(\widehat{\xi}_{n})$ are given by \eqref{DP-21}.

Similar to the case of simple poles, we also have the trace formulae for the case of double poles as
\begin{equation}\label{DP-17}
s_{11}(z)=e^{s(z)}s_{0}(z)~~\mbox{for}~~z\in D_{+}^{f},~~s_{22}(z)=e^{-s(z)}/s_{0}(z)~~\mbox{for}~~z\in D_{-}^{f},
\end{equation}
where
\begin{equation}\label{DP-18}
s(z)=-\frac{1}{2\pi i}\int_{\Sigma^{f}}\frac{\log\left[1+\rho(\zeta)\rho^{*}(\zeta^{*})\right]}{\zeta-z}d\zeta,~~
s_{0}(z)=\prod_{n=1}^{N}\frac{(z-z_{n})^2\left(z+q_{0}^2/z^{*}_{n}\right)^2}{(z-z^{*}_{n})^2\left(z+q_{0}^2/z_{n}\right)^2}.
\end{equation}
In terms of the limit $z\rightarrow0$ of $s_{11}(z)$ in \eqref{DP-17}, one can generate the theta condition as
\begin{equation}\label{DP-19}
\mbox{arg}\left(\frac{q_{+}}{q_{-}}\right)=8\sum_{n=1}^{N}\mbox{arg}(z_{n})
+\int_{\Sigma^{f}}\frac{\log\left[1+\rho(\zeta)\rho^{*}(\zeta^{*})\right]}{\zeta-z}d\zeta.
\end{equation}
Particularly, for the reflectionless case, i.e, $\rho(z)=\hat{\rho}(z)=0$,  the following Theorem 4.2 holds.

\noindent
\textbf{Theorem 4.2.} The reflectionless potential with double poles of the Kudun-NLS equation \eqref{gtc-NLS} with NZBCs can be written as
\begin{equation}\label{DP-20}
u(x,t)=\left[q_{-}+i\frac{\det\left(
                          \begin{array}{cc}
                            H & v \\
                            w^{T} & 0 \\
                          \end{array}
                        \right)
}{\det H}\right]\frac{e^{-i\gamma}}{\epsilon},
\end{equation}
where
$H=\left(H^{(sj)}\right)_{2\times2},~~H^{(sj)}=\left(h_{kn}^{(sj)}\right)_{2N\times2N}$,
$h_{kn}^{(11)}=\widehat{C}_{n}(\xi_{k})\left(D_{n}+\frac{1}{\xi_{k}-\widehat{\xi}_{n}}\right)-\frac{iq_{-}}{\xi_{k}}\delta_{kn}$,
$h_{kn}^{(12)}=\widehat{C}_{n}(\xi_{k})$,
$h_{kn}^{(21)}=\frac{\widehat{C}_{n}}{\xi_{k}-\widehat{\xi}_{n}}(\xi_{k})\left(D_{n}+\frac{2}{\xi_{k}-\widehat{\xi}_{n}}\right)
-\frac{iq_{-}}{\xi_{k}^2}\delta_{kn}$,
$h_{kn}^{(22)}=\frac{\widehat{C}_{n}(\xi_{k})}{\xi_{k}-\xi_{n}}+\frac{iq_{-}q_{0}^2}{\xi_{k}^3}\delta_{kn}$,
$w_{n}^{(1)}=A_{-}[\widehat{\xi}_{n}]e^{2i\theta(\xi_{n})}\widehat{D}_{n}$,
$w_{n}^{(2)}=A_{-}[\widehat{\xi}_{n}]e^{2i\theta(\xi_{n})}$, $v_{n}^{(1)}=-\frac{q_{-}}{\xi_{n}}$,
$v_{n}^{(2)}=-\frac{q_{-}}{\xi^2_{n}}$.

Therefore, the trace formulae and theta condition reduce to
\begin{align}\label{DP-22}
&s_{11}(z)=\prod_{n=1}^{N}\frac{(z-z_{n})^2\left(z+q_{0}^2/z^{*}_{n}\right)^2}{(z-z^{*}_{n})^2\left(z+q_{0}^2/z_{n}\right)^2},~~z\in D_{+}^{f},\\
&s_{22}(z)=\prod_{n=1}^{N}\frac{(z-z^{*}_{n})^2\left(z+q_{0}^2/z_{n}\right)^2}{(z-z_{n})^2\left(z+q_{0}^2/z^{*}_{n}\right)^2},~~z\in D_{-}^{f},
\end{align}
and
\begin{equation}\label{DP-23}
\mbox{arg}\left(\frac{q_{+}}{q_{-}}\right)=8\sum_{n=1}^{N}\mbox{arg}(z_{n}),
\end{equation}
respectively.

In the following, the double-pole breather-breather solutions of the Kundu-NLS equation \eqref{gtc-NLS} with NZBCs are displayed in Figs.7 and 8,
which can help us understand the properties of the double-pole breather-breather solutions.
Specifically, when $q_{-}\rightarrow0$, we have the double-pole bright-bright solutions of the Kundu-NLS equation \eqref{gtc-NLS} (see Fig.7(d)).

$~~~~~~$
{\rotatebox{0}{\includegraphics[width=5.2cm,height=3.6cm,angle=0]{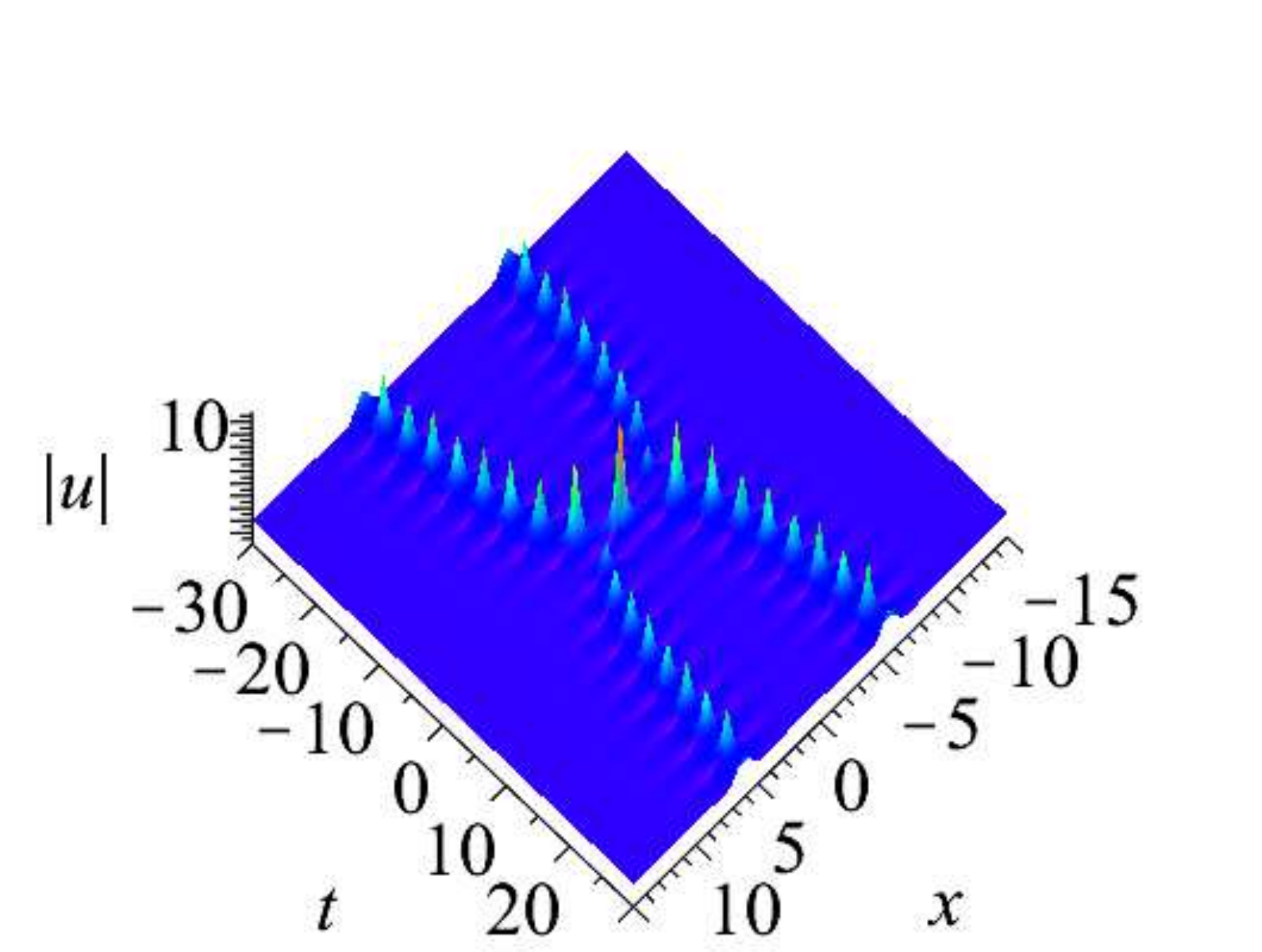}}}
~~~~~~~~~~~~~~~
{\rotatebox{0}{\includegraphics[width=5.2cm,height=3.6cm,angle=0]{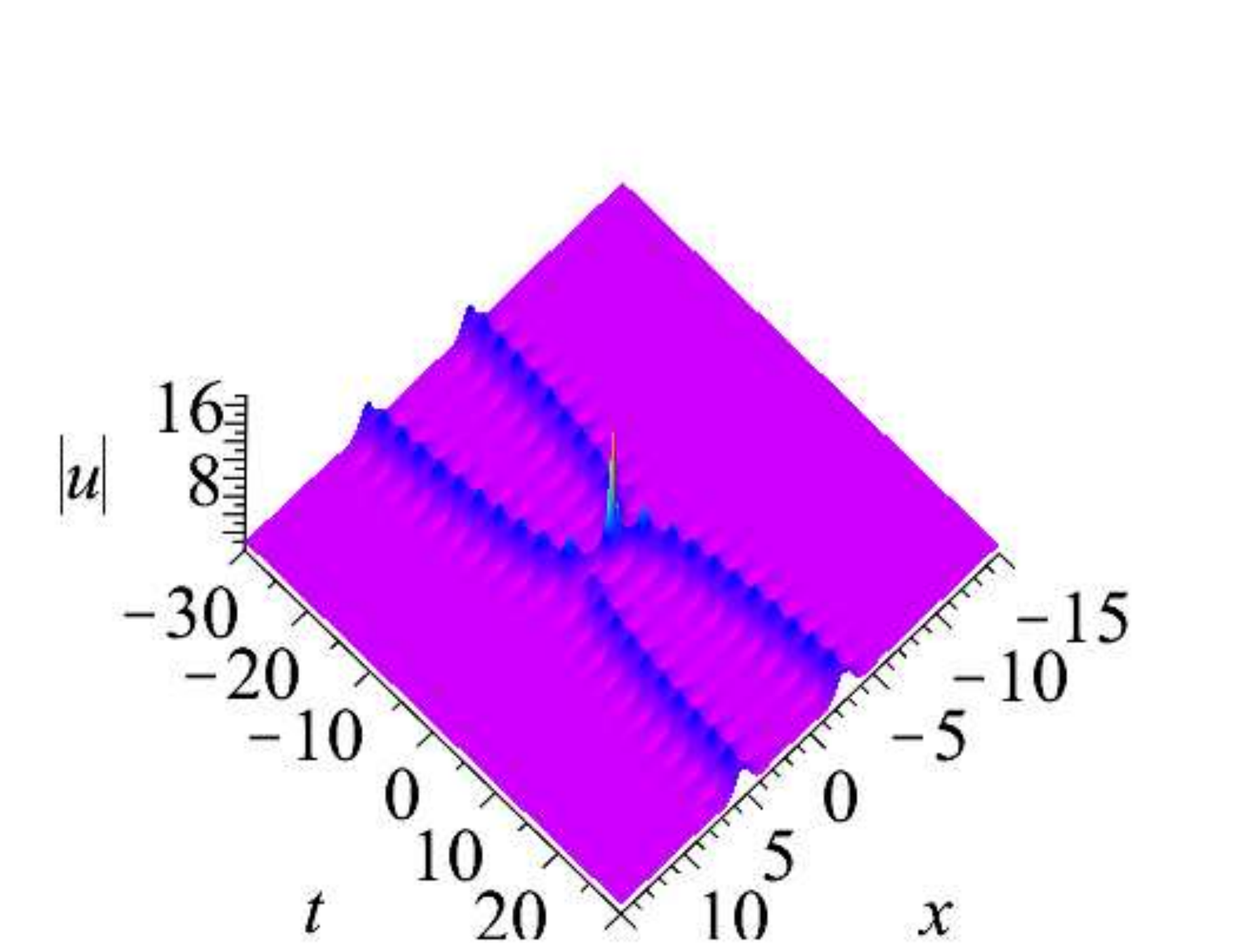}}}

$\qquad\qquad\qquad\qquad\textbf{(a)}
\qquad\qquad\qquad\qquad\qquad\qquad\qquad\qquad\qquad\textbf{(b)}
$\\

$~~~~~~$
{\rotatebox{0}{\includegraphics[width=5.2cm,height=3.6cm,angle=0]{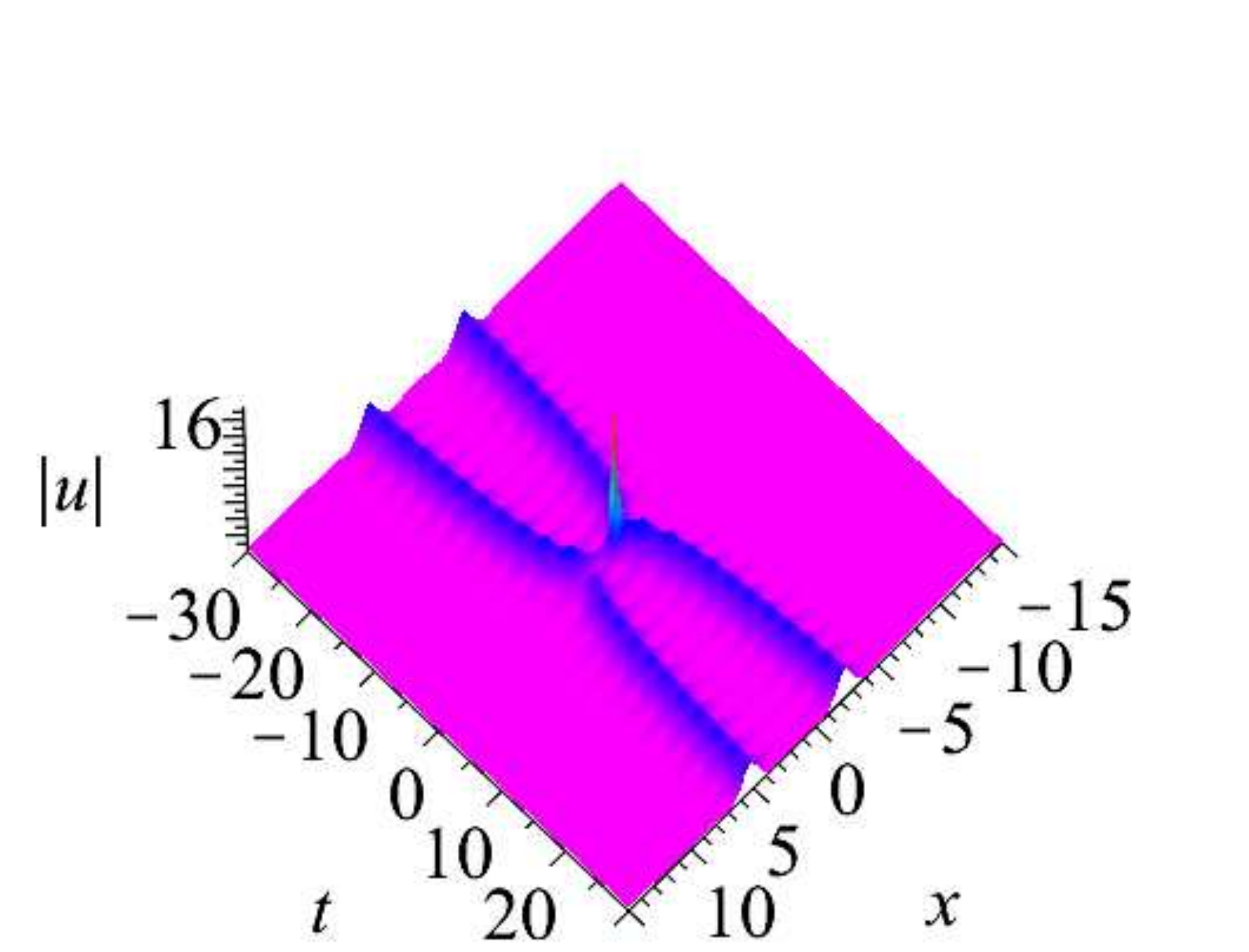}}}
~~~~~~~~~~~~~~~
{\rotatebox{0}{\includegraphics[width=5.2cm,height=3.6cm,angle=0]{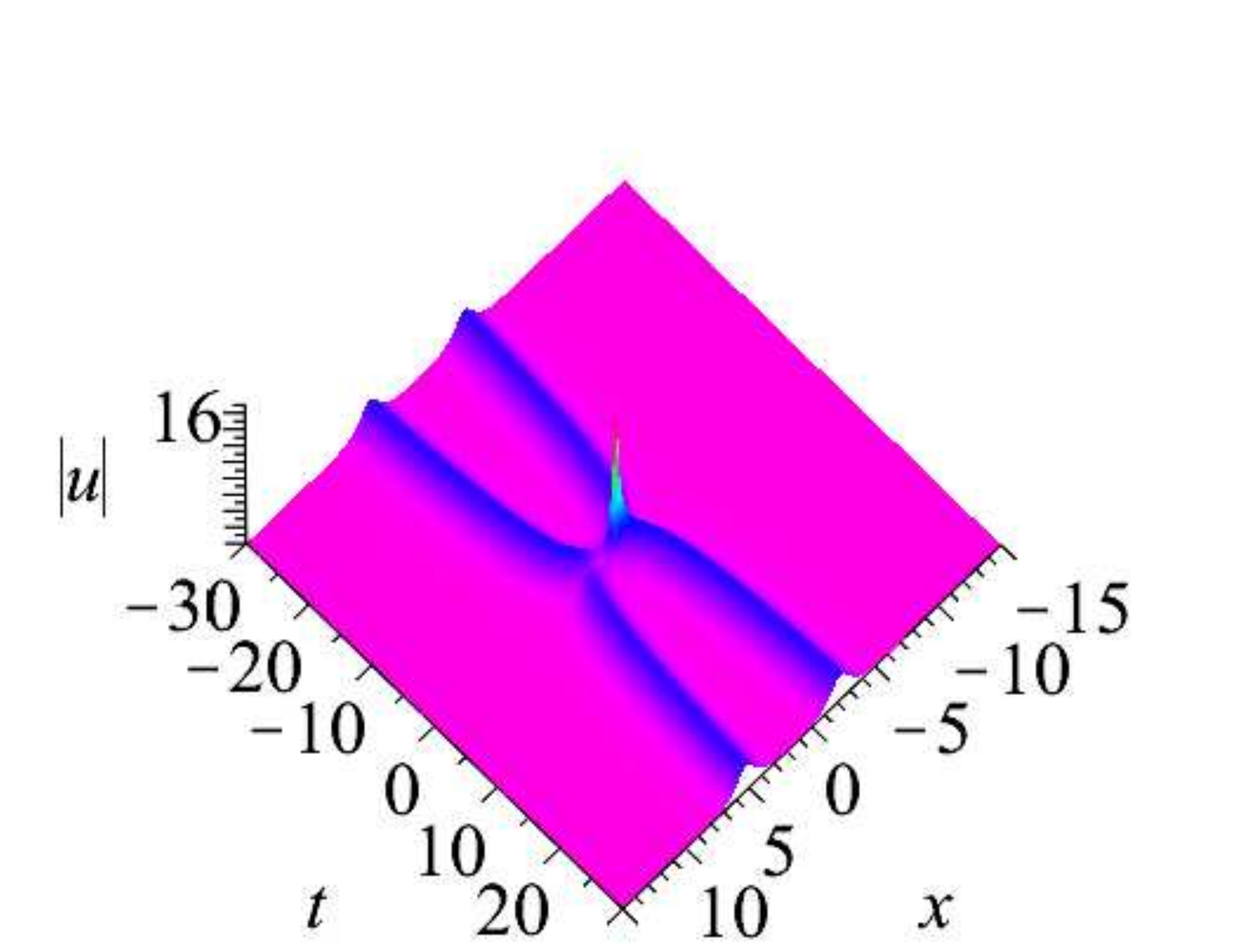}}}

$\qquad\qquad\qquad\qquad\textbf{(c)}
\qquad\qquad\qquad\qquad\qquad\qquad\qquad\qquad\qquad\textbf{(d)}
$\\
\noindent { \small \textbf{Figure 7.} (Color online) One-soliton solutions of the Kundu-NLS equation \eqref{gtc-NLS} with
$N=2,z_{1}=1.5i, A_{+}[z_{1}]=1,B_{+}[z_{1}]=1, \epsilon=0.5$:
$\textbf{(a)}$: breather-breather solution with $q_{-}=1$; $\textbf{(b)}$: breather-breather solution with $q_{-}=0.5$;
$\textbf{(c)}$: breather-breather solution with $q_{-}=0.2$;
$\textbf{(d)}$: bright-bright soliton solution with $q_{-}\rightarrow0$.\\}

$~~~~~~$
{\rotatebox{0}{\includegraphics[width=5.2cm,height=3.6cm,angle=0]{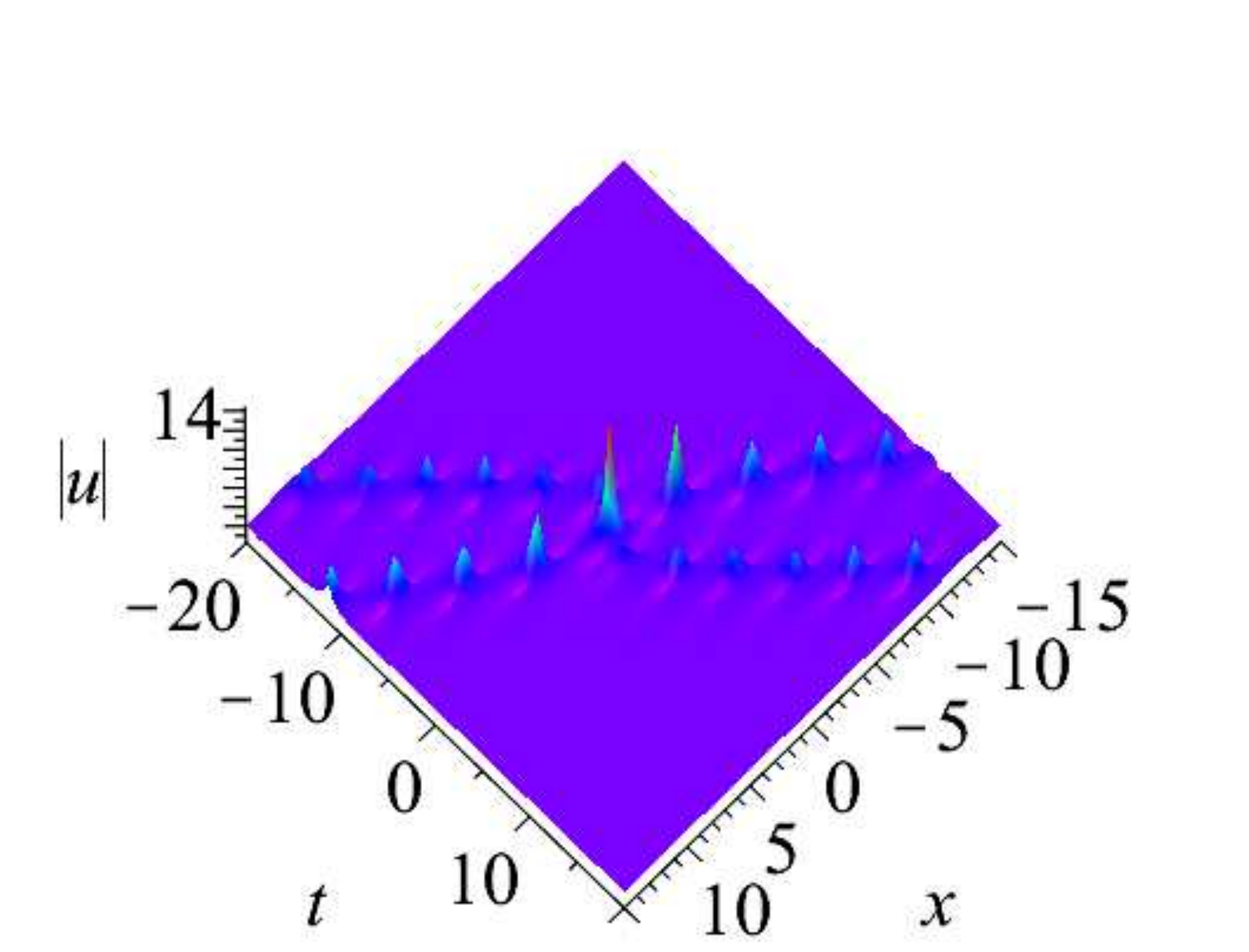}}}
~~~~~~~~~~~~~~~
{\rotatebox{0}{\includegraphics[width=5.2cm,height=3.6cm,angle=0]{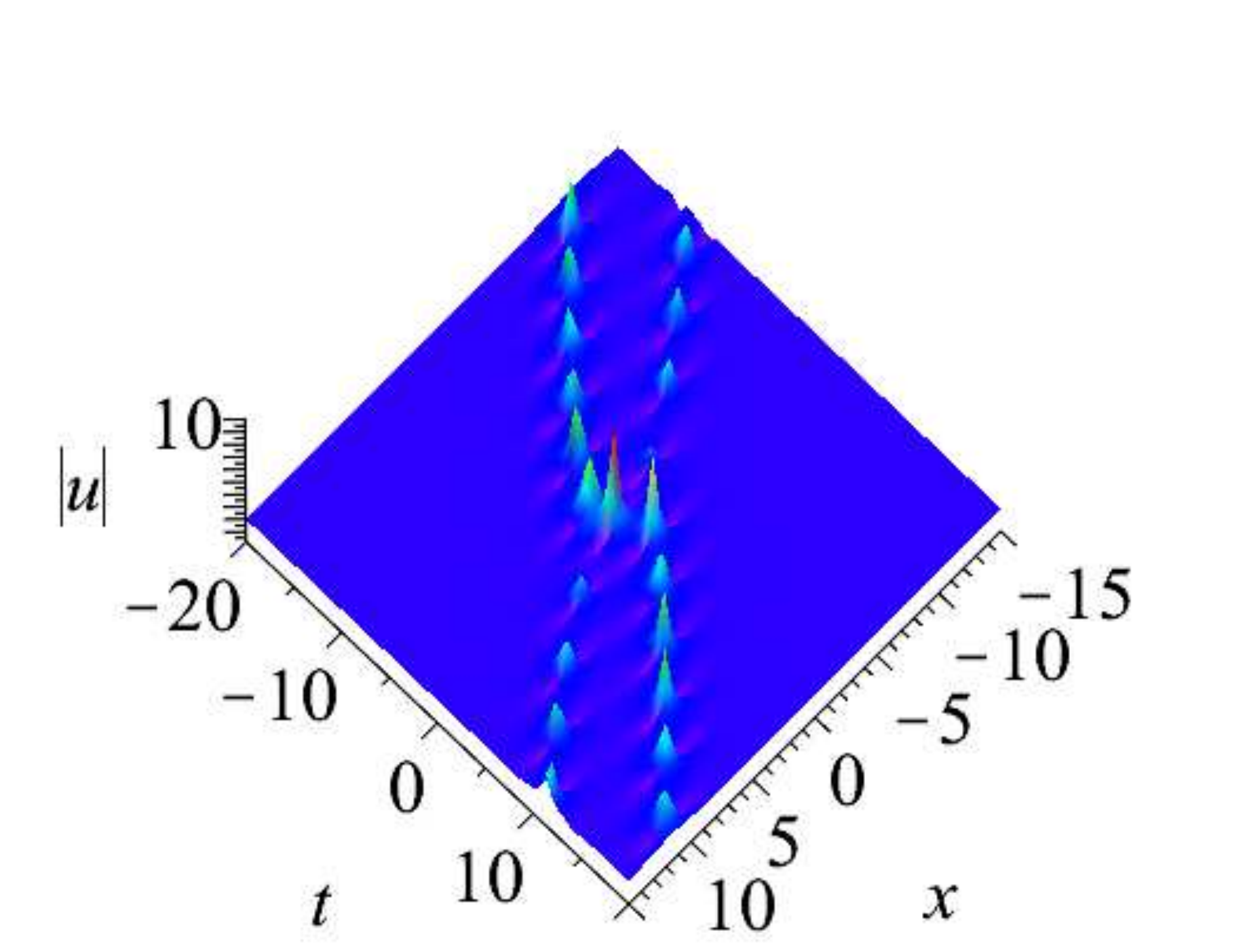}}}

$\qquad\qquad\qquad\qquad\textbf{(a)}
\qquad\qquad\qquad\qquad\qquad\qquad\qquad\qquad\qquad\textbf{(b)}
$\\

$~~~~~~$
{\rotatebox{0}{\includegraphics[width=5.2cm,height=3.6cm,angle=0]{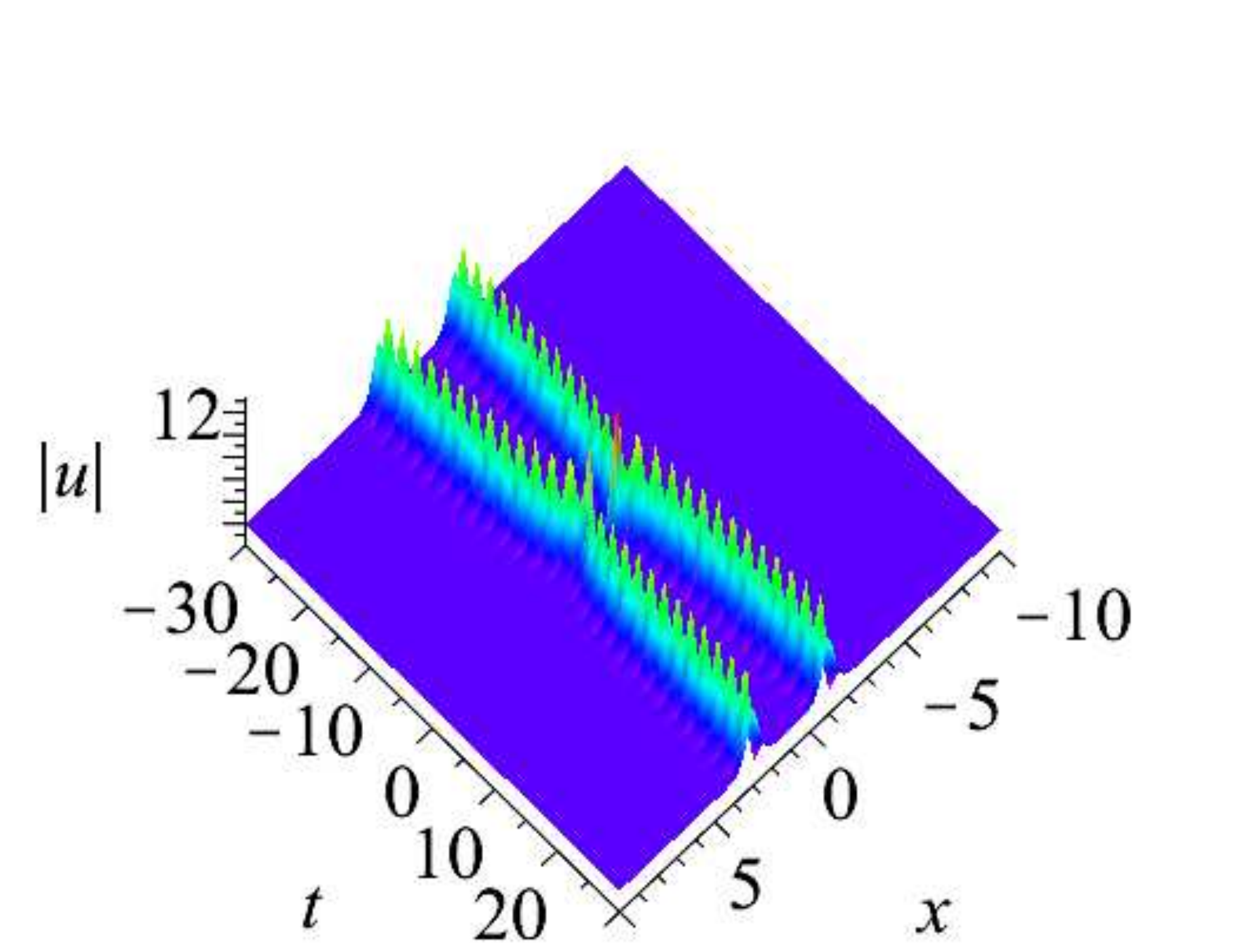}}}
~~~~~~~~~~~~~~~
{\rotatebox{0}{\includegraphics[width=5.2cm,height=3.6cm,angle=0]{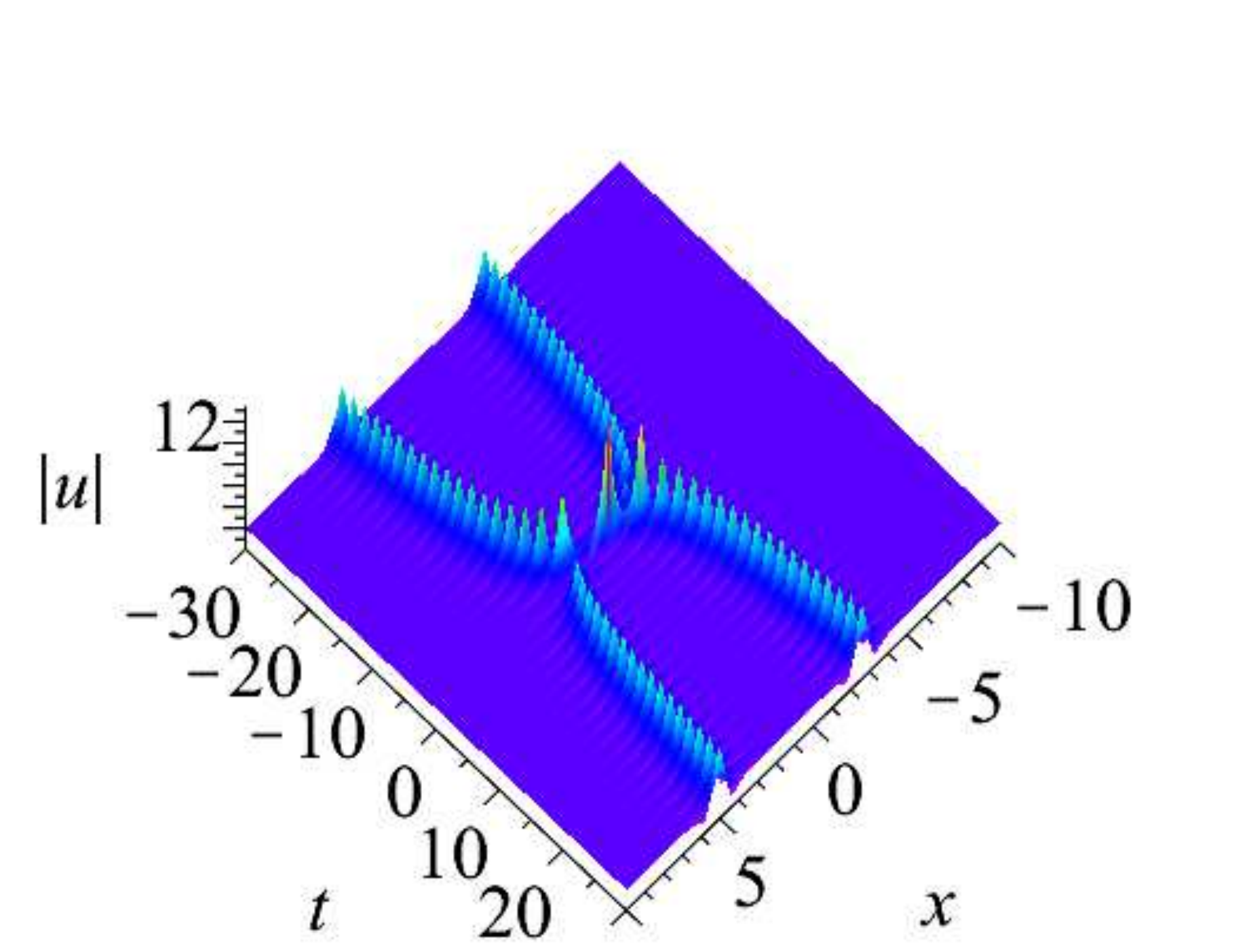}}}

$\qquad\qquad\qquad\qquad\textbf{(c)}
\qquad\qquad\qquad\qquad\qquad\qquad\qquad\qquad\qquad\textbf{(d)}
$\\
\noindent { \small \textbf{Figure 8.} (Color online) One-soliton solutions of the Kundu-NLS equation \eqref{gtc-NLS} with
$N=2,q=q, A_{+}[z_{1}]=1,B_{+}[z_{1}]=1, \epsilon=0.5$:
$\textbf{(a)}$: $z_{1}=0.2+1.5i$; $\textbf{(b)}$: $z_{1}=-0.2+1.5i$;
$\textbf{(c)}$: $z_{1}=2i$
$\textbf{(d)}$: $z_{1}=4i$.\\}

\section{Conclusion and Discussion}
In summary, we have discussed the IST and soliton solutions for the Kundu-NLS equation \eqref{gtc-NLS} with NZBCs starting from its Lax pair.
Then we investigate the Kundu-NLS equation \eqref{gtc-NLS} with NZBCs such that its simple-pole and double-pole solutions are found
via solving the matrix RHP with the reflectionless potentials.
Then some representative solitons are well constructed.
Moreover, in order to help the readers understand the solutions
better, Figs.2-8 of the breather waves, bright soliton,  breather-breather waves and bright-bright solitons, respectively, are plotted
by seeking appropriate parameters.
More importantly,
Our results would be of much importance in enriching rogue wave phenomena in nonlinear wave fields.
The work shows that the IST provides a direct
and powerful mathematical tool to construct exact solution of other nonlinear wave equations, which can be suitable to
analyze other models in mathematical physics and engineering.

\section*{Acknowledgements}
\hspace{0.3cm}This work is supported by the National Natural Science Foundation of China under Grant No.11871180.

%\end{CJK*}
\end{document}